\documentclass[iicol, referee, pdflatex, sn-mathphys-num]{sn-jnl}

\usepackage{graphicx}%
\usepackage{multirow}%
\usepackage{amsmath,amssymb,amsfonts}%
\usepackage{amsthm}%
\usepackage{mathrsfs}%
\usepackage[title]{appendix}%
\usepackage{xcolor}%
\usepackage{textcomp}%
\usepackage{manyfoot}%
\usepackage{booktabs}%
\usepackage{algorithm}%
\usepackage{algorithmicx}%
\usepackage{algpseudocode}%
\usepackage{listings}%

\usepackage{siunitx} 
\usepackage{upgreek} 
\usepackage{bm} 
\usepackage{tabularx}

\onecolumn 

\theoremstyle{thmstyleone}%
\theoremstyle{thmstyletwo}%

\theoremstyle{thmstylethree}%

\def\paperTitle{Speckle-based X-ray microtomography via preconditioned Wirtinger flow
}
\begin{document}
\title[Article Title]{\paperTitle}

\author*[1,2,4]{\fnm{KyeoReh} \sur{Lee}}\email{lee.kyeo@gmail.com}
\equalcont

\author[1,2]{\fnm{Herve} \sur{Hugonnet}}\email{grougrou@kaist.ac.kr}
\equalcont{These authors contributed equally to this work.}

\author[3]{\fnm{Jae‑Hong} \sur{Lim}}\email{limjh@postech.ac.kr}

\author*[1,2]{\fnm{YongKeun} \sur{Park}}\email{yk.park@kaist.ac.kr}

\affil[1]{\orgdiv{Department of Physics}, \orgname{Korea Advanced Institute of Science and Technology}, \orgaddress{\city{Daejeon}, \postcode{34141}, \country{Republic of Korea}}}

\affil[2]{\orgdiv{KAIST Institute for Health Science and Technology}, \orgname{Korea Advanced Institute of Science and Technology}, \orgaddress{\city{Daejeon}, \postcode{34141}, \country{Republic of Korea}}}

\affil[3]{\orgdiv{Pohang Accelerator Laboratory}, \orgname{Pohang University of Science and Technology}, \orgaddress{\city{Pohang}, \postcode{37637}, \country{Republic of Korea}}}

\affil*[4]{Current address: \orgdiv{Department of Applied Physics}, \orgname{Yale University}, \orgaddress{\city{New Haven}, \postcode{06520}, \state{Connecticut}, \country{USA}}}

\abstract{
Three-dimensional quantitative phase imaging has been extensively studied in X-ray microtomography to improve the sensitivity and specificity of measurements, especially for low atomic number materials. However, obtaining quantitative phase images typically requires additional measurements or assumptions, which significantly limits the practical applicability. Here, we present preconditioned Wirtinger flow (PWF) to realize an assumption-free, single-shot, quantitative X-ray microtomography. Accurate phase retrieval is demonstrated using a specialized gradient-based algorithm with an accurate physical model. Partial coherence of the source is taken into account, extending the potential applications to bench-top sources. Improved accuracy and spatial resolution over conventional speckle tracking methods are experimentally demonstrated. The various samples are explored to demonstrate the robustness and versatility of PWF.
}
\keywords{X-ray imaging, Computed tomography, Speckle tracking, Partially coherent imaging, Phase retrieval}
\maketitle

\section{Introduction}\label{SecIntroduction}

Due to the inherently low absorption contrast of low-atomic-number materials, phase-contrast techniques have long been of interest in the field of hard X-ray imaging \cite{momose2005recent}.

The first and arguably most popular method is to defocus the sample, which highlights the sample edges \cite{snigirev1995possibilities, cloetens1996phase, pogany1997contrast}. While such edge detection helps in understanding the internal structures of the sample, directly retrieving the sample phase shift is challenging without prior knowledge of the sample \cite{burvall2011phase}. To alleviate this limitation, multiple acquisitions at different sample-to-detector distances have been used \cite{nugent1996quantitative, cloetens1999holotomography}. However, translating either the sample or the detector along the beam direction often introduces additional technical issues such as image registration \cite{zanette2013holotomography}.

Another popular approach is to introduce a modulator between the sample and the detector. A representative example is the Hartmann sensor, which uses a pinhole array as a modulator \cite{mercere2003hartmann}. The displacement in each pinhole projection pattern provides a local phase gradient map, and the quantitative phase can be obtained by numerical integration \cite{salas2005wave}. Although it remains popular in beam characterization \cite{keitel2016hartmann}, it has rarely been used for phase imaging due to the trade-off between image resolution and crosstalk between subregions.

Grating shearing interferometry is often preferred for phase imaging applications \cite{david2002differential, weitkamp2005x}. The shear between the grating diffraction orders allows the local phase gradient to be obtained from the interference without sacrificing spatial resolution. In optical microscopy with sufficient magnification, it is possible to isolate the interference term from a single image by using off-axis interferometry \cite{bon2009quadriwave}. In contrast, projection X-ray imaging typically uses an analyzer grating to demodulate the interferometric fringe, requiring multiple acquisitions to isolate the interference term \cite{weitkamp2005x}. The use of physical interference ensures the credibility of the measured phase gradient, which is a distinct advantage \cite{zanette2013holotomography}. Despite its usefulness, the requirement for high-aspect-ratio X-ray gratings significantly increases the technical difficulty of reproduction, especially for the two-dimensional gratings needed for unambiguous phase determination \cite{kottler2007two, zanette2010two}.

Recently, speckle-tracking methods have been proposed that introduce a diffuser as a modulator \cite{morgan2012x, berujon2012two, berujon2012x, wu2024multiplexed}. The use of ordinary paper or sandpaper instead of special pinhole arrays or gratings significantly reduces the technical difficulty. Similar to the Hartmann sensor, the local phase gradient is derived from the displacements of subregions, which in this case are random speckle patterns. Due to the sharply peaked autocorrelation function of the speckle, it is possible to uniquely determine the local displacements without crosstalk \cite{goodman2007speckle}. Since accurate local displacement estimation is crucial for robust phase retrieval, several phase retrieval algorithms have been developed. One popular approach is direct cross correlation calculation based on subregion partitioning, similar to that of Hartmann sensors \cite{zanette2014speckle, berujon2015near, zdora2017x}. Another popular approach is to use the transport intensity equation (TIE) concept, originally introduced in defocusing methods \cite{paganin2018single, pavlov2020x, quenot2021implicit, alloo2023m}. Although both approaches work reasonably well in practical situations \cite{zdora2020x, savatovic2023multi, magnin2023dark}, they rely on assumptions or approximations that simplify the physical imaging model. We summarized the assumptions, required measurement of speckle-tracking methods in Table~\ref{tab:STMcomparative}. As a result, it is difficult to determine whether they fully capture the sample field information encoded in the speckle patterns.

Fortunately, speckle-based imaging has already been studied in various disciplines for both coherent \cite{stockmar2013near, lee2016exploiting, zhang2016phase, mcdermott2018near, levitan2020single, oh2022single} and incoherent imaging systems \cite{labeyrie1970attainment, bertolotti2012non, katz2014non, antipa2017diffusercam}. In particular, the phase retrieval from a single speckle pattern has been extensively explored in the fields of computational imaging \cite{lee2016exploiting, zhang2016phase, levitan2020single, oh2022single} and applied mathematics \cite{candes2013phaselift, candes2015phase, wang2017solving, goldstein2018phasemax}. By exploiting the mathematical properties of random variables, robust phase retrieval from a single speckle pattern has been consistently reported without additional approximations or assumptions \cite{lee2016exploiting, candes2015phase}. In contrast to speckle-tracking methods, significant sample-induced speckle decorrelation is usually assumed, and the phase information is retrieved from the random self-interference of the incident field \cite{oh2022single}. Interestingly, these multidisciplinary works share common requirements regarding the number of measured speckle grains relative to the number of unknown variables. This is referred to as the resolution ratio \cite{levitan2020single}, the oversampling rate \cite{candes2013phaselift}, or the oversampling ratio \cite{lee2016exploiting, goldstein2018phasemax}.

\begin{figure}[t]
\centering
\includegraphics[width=1.0\textwidth]{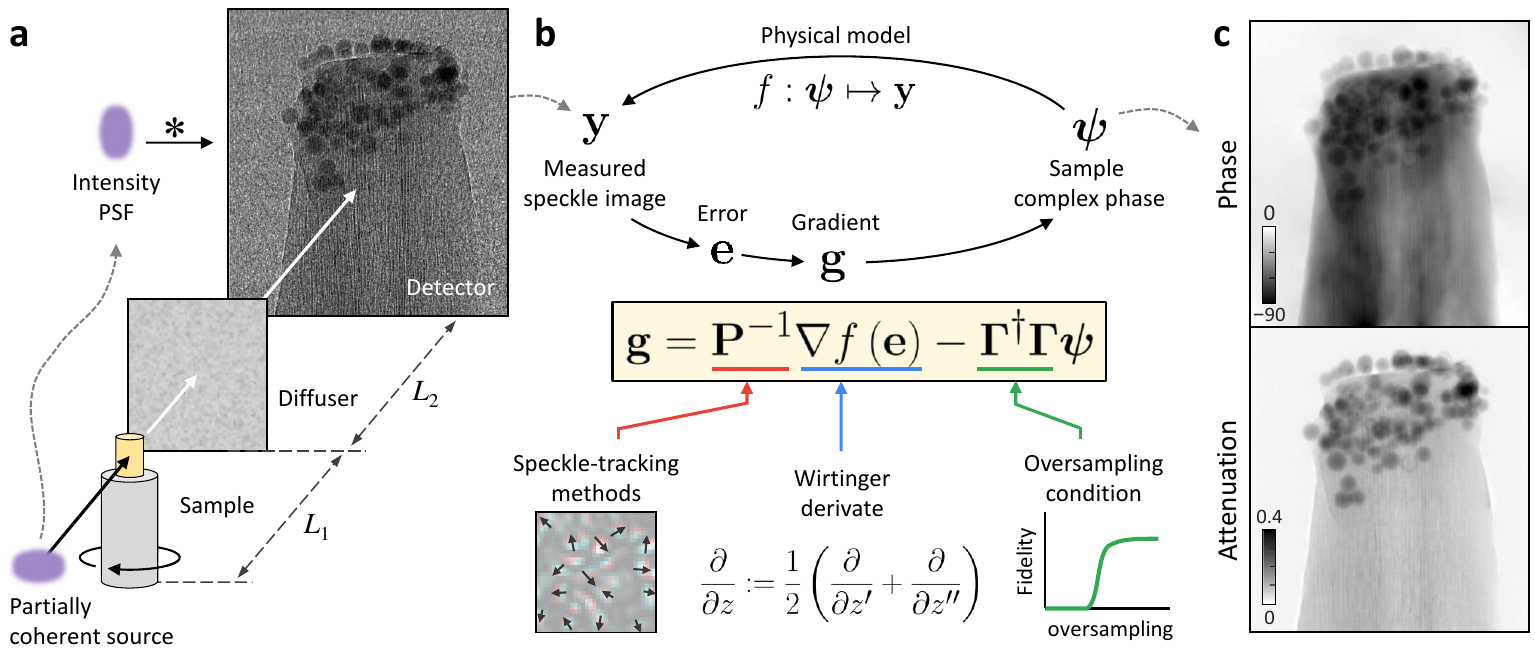}
\caption{
\textbf{Speckle–based X-ray microtomography with a partially coherent source.}
\textbf{a}, Schematic of the experimental setup. Partially coherent source provides an additional blurring effect on the intensity image. The $*$ symbol indicates image convolution. $L_1=\qty{3}{\mm}$ and $L_2=\qty{20}{\mm}$ are used throughout the paper. An experimentally measured speckle image of the toothpick with glass beads sample is presented on the detector plane.
\textbf{b}, Schematic of preconditioned Wirtinger Flow (PWF). The vectors are defined as $\bm{\psi}:=\left[\psi_r \right]_{1\leq r \leq m}$, where $m$ is the number of pixels in the image. Although the variables and operators are represented in vector-matrix form here for visualization purposes, the actual algorithm is performed based on element-wise calculations and Fourier transforms (see Algorithm~\ref{algorithm:PWF}). $z=z'+iz''$ is a complex number.
\textbf{c}, Reconstructed phase and attenuation from the measured speckle pattern (see Fig.~\ref{fig2} for details)
}
\label{fig1}
\end{figure}

Despite significant discrepancies in experimental conditions, the mathematical principles and algorithms of speckle-based imaging should prevail for conventional speckle-based X-ray microtomography (Fig.~\ref{fig1}a), since both methods decode the sample field information from measured speckle patterns.
Then, is it possible to find a phase retrieval algorithm that incorporates all the lessons learned from the aforementioned speckle-related methods? In this paper, we present the preconditioned Wirtinger flow (PWF) as the solution (Fig.~\ref{fig1}b). Exploiting the Wirtinger flow as a backbone \cite{candes2015phase}, we introduce three core ideas: a reliable physical model that includes the decoherence effect of the partially coherent X-ray source (Fig.~\ref{fig1}a); an preconditioner inspired from speckle-tracking methods ($\mathbf{P}^{-1}$ in Fig.~\ref{fig1}b); and a Tikhonov regularizer to satisfy the oversampling ratio criterion ($\mathbf{\Gamma}^\dagger\mathbf{\Gamma}$ in Fig.~\ref{fig1}b). We successfully demonstrate the effectiveness of PWF in experiments (Fig.~\ref{fig2}). From the corresponding tomogram results (Fig.~\ref{fig3}), we find that PWF, with a single speckle measurement, outperforms existing speckle-tracking methods, which uses 12 different speckle measurements, in both refractive index accuracy and resolution. We also demonstrate the versatility of PWF by reconstructing phase tomograms in various samples (Fig.~\ref{fig4}).

\section{Results}\label{sec:Results}
\subsection{Preconditioned Wirtinger flow}\label{subsec:PWF}
We propose PWF to retrieve the complex sample field from the conventional speckle-based X-ray microtomography (Fig.~\ref{fig1}a). The reconstruction requires a single sample speckle pattern ($y_r$) with a reference speckle pattern that is acquired without the sample. The X-ray wavelength (\qty{0.124}{\nm}, \qty{10}{\keV}), and the distances before and after the diffuser ($L_1 = \qty{3}{\mm}$ and $L_2 = \qty{20}{\mm}$) are known constants. The detailed PWF algorithm is described in Algorithm~\ref{algorithm:PWF}. Throughout the paper, we use vector indices $r$ and $k$ for real and reciprocal (or Fourier) space, respectively.

Instead of the sample field ($x_r$), we directly reconstruct the complex phase ($\psi_r = \log x_r$) to avoid potential instability caused by phase wrapping \cite{wittwer2022phase}. Note that the real and imaginary parts of $\psi_r$ represent the attenuation (with a negative sign) and the unwrapped phase of the sample, respectively. Now, the physics-based forward model ($f:\psi_r \mapsto y_r$) can be defined as follows
\begin{equation} 
f(\psi_r)= 
        \left| \mathcal{P}_2 
            \left\{
                t_r \mathcal{P}_1 \left\{ e^{\psi_r} \right\}
            \right\}
        \right|^2\ast\mathrm{IPSF}_r,
\label{eq:forwardModel}
\end{equation} 
where $\mathcal{P}_{1,2}\left\{\cdot\right\}$ are the free-space propagation operators for the distances $L_{1,2}$ based on angular-spectrum method, $t_r$ is the diffuser transmission function, $\mathrm{IPSF}_r$ is the intensity point spread function, and $*$ is the image convolution operator (Fig.~\ref{fig1}b).

We derive the diffuser transmission function ($t_r$) from the reference speckle. A robust sample field reconstruction is found even with the phase ambiguity of the reference speckle (see Methods). The effect of partially coherent source is applied by convoluting an intensity point spread function ($\mathrm{IPSF}_r$) as described in Fig.~\ref{fig1}a; or equivalently, by applying the intensity optical transfer function ($\mathrm{IOTF}_k$) as a window function. We experimentally obtain $\mathrm{IOTF}_k$ from the power spectral density of the reference speckle, based on the Siegert relation (see Methods). Retrieved spatial horizontal and vertical coherence lengths ($x_\mathrm{coh}$ and $y_\mathrm{coh}$) are \qtylist{3.47; 4.31}{\um}, respectively (Fig.~\ref{figS:speckle}). It is noteworthy that $\mathrm{IOTF}_k$ here is identical to the ``coherence envelope'' in Ref.~\cite{pogany1997contrast}.

At this point, the phase retrieval problem can be understood as finding the $\psi_r$ that best reproduces the measured speckle pattern based on the defined physical model,
\begin{equation} 
\underset{\psi_r}{\operatorname{minimize}}\ \sum_r\left(y_r-f(\psi_r)\right)^2.
\label{eq:lossFunction}
\end{equation}
PWF is a gradient-based solver for Eq.~\ref{eq:lossFunction}, that seeks the solution by error backpropagation $\nabla f(e_r)$, where $e_r=y_r-f(\psi_r)$ (Fig.~\ref{fig1}b). We use the Wirtinger derivative to compute the gradient of non-holomorphic complex functions to perform this minimization. As the measured intensity images is already a non-holomorphic function, most iterative algorithms have employed the Wirtinger derivative to backpropagate the error, either explicitly or implicitly. For instance, ptychographic iterations  can be regarded as a variant that utilizes the Wirtinger derivative \cite{thibault2008high, maiden2009improved}, albeit with a modified loss function \cite{wang2017solving}.

Preconditioning is a popular technique for improving the convergence properties of iterative solvers, especially for ill-conditioned problems \cite{saad2003iterative}. For a linear problem $\mathbf{Ax}=\mathbf{b}$, preconditioning can be understood as the compensation of the inhomogeneous singular value distribution of $\mathbf{A}$, effectively making its gradient direction $\mathbf{A}^\dagger$ more parallel to $\mathbf{A}^{-1}$. For complex nonlinear problems, however, the error propagation properties are often obscure, making it difficult to define an appropriate preconditioner. 

Here, we determine the preconditioner with the help of speckle-tracking methods. Based on the working principle of speckle-tracking methods and their robustness, we understand that our imaging system is well-conditioned for the sample phase gradient rather than the phase itself. Since the phase gradient is equivalent to the filtered phase with linear spatial frequency ramps, the phase retrieval iteration is expected to be highly ill-conditioned, leading to large convergence rate discrepancy between high and low frequency (Fig.~\ref{figS:noPrecon}). Accordingly, we introduce an inverse quadratic preconditioning filter on the imaginary part of $\psi_r$ to invert the linear spatial frequency ramps and improve the convergence properties (see Methods).

The regularizer imposes the oversampling criterion in speckle-based imaging, which requires a sufficient oversampling ratio to ensure unique determination of the phase and amplitude in field reconstruction \cite{candes2015phase, lee2016exploiting}. The oversampling ratio is defined as $\gamma = M/N$, where $M$ and $N$ are the number of measured and reconstructed spatial modes \cite{oh2022single, lee2023direct}. We compute the space-bandwidth product (SBP) to estimate $M$ and $N$, which then determine the regularization window for a given $\gamma$ (see Methods). 

The oversampling ratio ($\gamma$) and the regularization parameter ($\alpha$) are determined heuristically. Since they usually depend strongly on the experimental signal-to-noise ratio (SNR), these parameters would depend on the imaging setup. Comparative results with different $\gamma$ and $\alpha$ are shown in Figs.~\ref{figS:gamma} and \ref{figS:alpha}, respectively. Although different parameters would give different results, we did not find a strong dependence on the parameters between samples. Therefore, a fixed oversampling ratio ($\gamma = 1$) and regularization parameter ($\alpha = 0.1+0.01i$) are used throughout the paper. The real and imaginary parts of $\alpha$ are applied separately to the attenuation and phase of the sample field.
The detailed calculations for the physical model, preconditioner, and regularizer are presented in Algorithm~\ref{algorithm:functions}. The MATLAB code has been uploaded online to facilitate reproduction (see Data availability). To further accelerate the convergence, Nesterov's accelerated gradient method was employed \cite{nesterov1983method} (Algorithm \ref{algorithm:PWF}). A fixed step size ($\eta$ = 1) is used throughout the paper.

\subsection{Field reconstruction results}\label{subsec:fieldRecon}
\begin{figure}[t]
\centering
\includegraphics[width=1.0\textwidth]{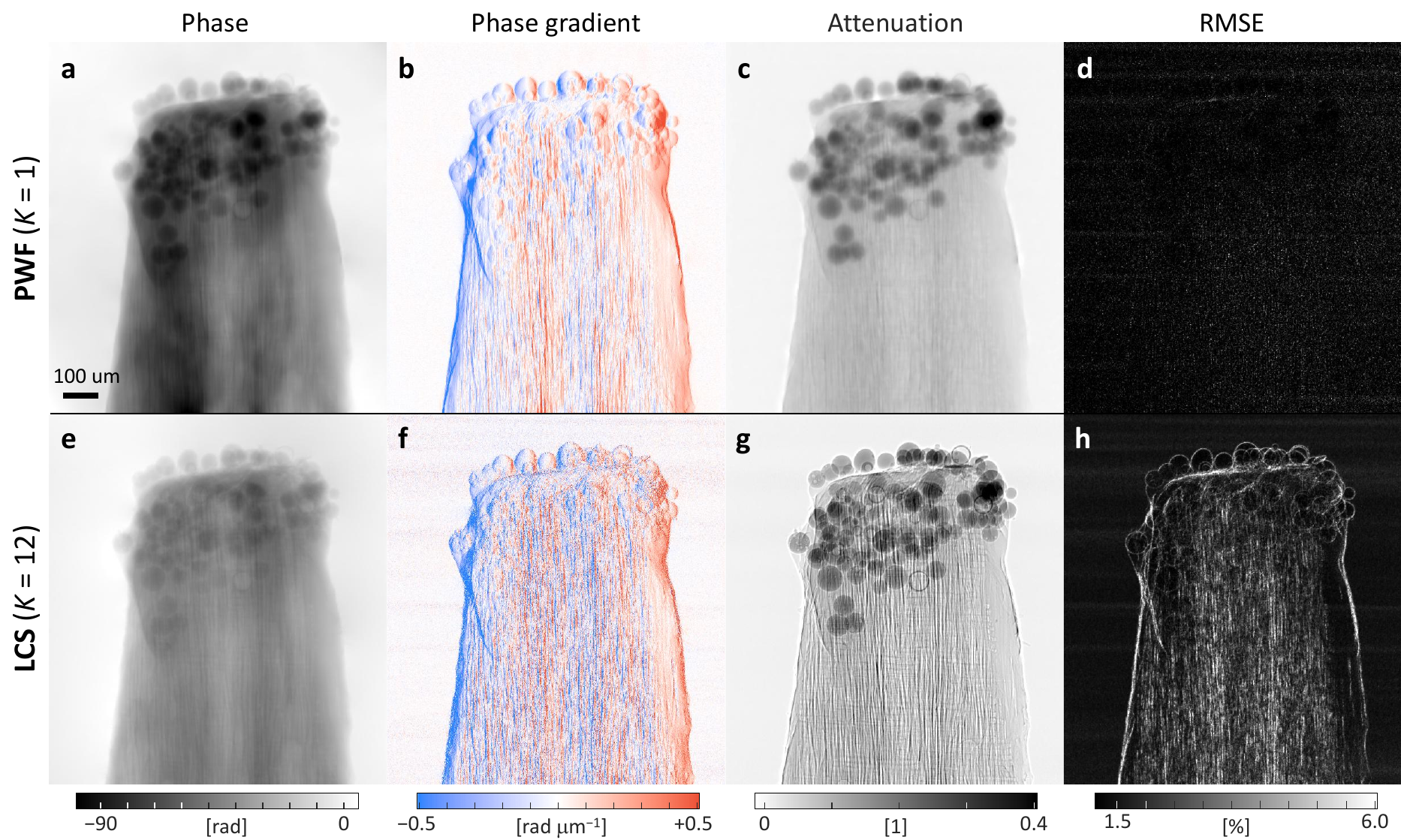}
\caption{
\textbf{Field reconstruction results of a toothpick with glass beads.}
\textbf{a--d}, Reconstructed phase (\textbf{a}), horizontal phase gradient (\textbf{b}), and attenuation (\textbf{c}), and the remaining root-mean-square error (RMSE, \textbf{d}) from PWF with a single measurement ($K$ = 1). \textbf{e--h}, The reconstructed counterparts from LCS with 12 measurements at different diffuser positions ($K$ = 12) \cite{quenot2021implicit}.
}\label{fig2}
\end{figure}

The experiments are performed at the X-ray microtomography beamline (6C, PLS-II) using a \qty{10}{\keV} partially coherent X-ray from a wiggler source (see Methods). A toothpick with glass beads is chosen as the first sample (see Methods). The measured raw speckle image is shown in Fig.~\ref{fig1}a and the corresponding reconstruction results are shown in Fig.~\ref{fig2}.
 The raw speckle and reconstructed images have the same number of pixels (\numproduct{2560 x 2160}). The PWF algorithm takes \qtyrange{5}{6}{\s} to converge on a personal computer (see Methods).

Along with PWF, we employ the low coherence system (LCS) implicit speckle-tracking method for a comparison \cite{quenot2021implicit}. It is the best-performing and assumption-free speckle-tracking method among the algorithms we tested \cite{zdora2017x, paganin2018single, quenot2021implicit} (Fig.~\ref{figS:STMcomparison}). Unlike PWF, which requires a single measurement ($K$ = 1), speckle-tracking methods typically require multiple measurements for reliable phase reconstruction \cite{zdora2017x,quenot2021implicit}. Although a few speckle-tracking methods use a single speckle, they usually rely on strong assumptions such as no attenuation \cite{paganin2018single} and single material \cite{pavlov2020single}. Accordingly, the acquisition process is repeated at 12 different diffuser positions ($K$ = 12). 

Despite fewer measurements, we find that PWF consistently outperforms LCS (Fig.~\ref{fig2}). Significantly greater phase shifts are obtained (Figs.~\ref{fig2}a and \ref{fig2}e), where phase shift here refers to the magnitude of phase values. Note that phase shifts greater than $30\pi$ are obtained, indicating that the sample is optically very thick. This result is meaningful in itself, as robust field reconstruction of such a thick sample is usually difficult due to the complicated loss function landscape induced by severe phase wrapping \cite{maiden2017further}. Interestingly, perhaps due to the existence of the preconditioner, we find that the PWF also exhibits lower sensitivity to slow phase variations as in phase-gradient sensing techniques \cite{choi2017compensation} (see Fig.~\ref{figS:PWFsimul_phase}). Fortunately, it is known that such low-frequency inaccuracy is largely mitigated by averaging measurements taken at different projection angles during the tomographic reconstruction \cite{pfeiffer2007hard}. Since our goal is to obtain a quantitative 3D phase image, the fidelity of the measured phase values is discussed in the next section alongside the tomographic reconstruction results (Fig.~\ref{fig3}e). Less noisy phase gradient values are also observed, especially for the toothpick interior containing fine structures (Fig.~\ref{fig2}b and \ref{fig2}f). Although the PWF algorithm directly provides an integrated phase image, the phase gradient is calculated for direct comparison with LCS to eliminate potential effects of the integration method used.

Notably reduced edge-enhancement effect is observed in the attenuation result of PWF (Fig.~\ref{fig2}c and \ref{fig2}g). Since the enhancement at the edges originates from the abrupt local sample phase modulation \cite{pogany1997contrast}, the edge-enhancement artifacts in the attenuation image indicates an inefficient decoupling of phase shift and attenuation effects. We find that PWF successfully decouples the sample attenuation from the phase contribution despite the non-negligible propagation distance between the sample and the detector. This result strengthens the credibility of PWF in both phase and attenuation images.

Perhaps the most straightforward metric is the residual error, which directly indicates the discrepancy between the reconstructed results and the measurements. To quantify the residual error, we compute estimated speckle patterns based on the model and then obtain the root-mean-square error (RMSE) with respect to the measured speckle patterns (see Methods). Unlike PWF, which shows a uniform and noise level RMSE (Fig.~\ref{fig2}d), LCS shows an RMSE image that is similar to a ``dark-field'' image (Fig.~\ref{fig2}h) \cite{zanette2014speckle, pavlov2020x, magnin2023dark}. This similarity makes sense as speckle-tracking methods often obtain the dark-field images by introducing an additional term to explain the part that cannot be interpreted by the original speckle displacement model \cite{pavlov2020x, magnin2023dark}. An important question arises here: if the RMSE corresponds to the dark-field signal, why is it absent from PWF? 

The answer lies in the physical model of PWF. Note that the dark-field signal in projection X-ray imaging represents the fraction of sample diffraction that exceeds the coherence volume \cite{pfeiffer2008hard}. In other words, the dark-field is an alternative representation of the high-angle sample diffraction that cannot be acquired due to sample-induced decoherence. We find that PWF can incorporate some of the dark-field signal into the reconstructed sample field as high spatial frequency components. Despite the speckle blur, PWF is able to pick up the displacement as the forward model includes the decoherence effect (Eq.~\ref{eq:forwardModel}). This is analogous to incoherent speckle-based imaging methods that can reconstruct speckle displacements even from completely blurred speckle patterns \cite{labeyrie1970attainment, bertolotti2012non, katz2014non, antipa2017diffusercam}. Such high-angle diffraction signal reconstruction capability of PWF suggests an enhancement in resolution, which is discussed in the next section (Fig.~\ref{fig3}f).

\subsection{Tomogram results}\label{subsec:tomoRecon}
\begin{figure}[t]
\centering
\includegraphics[width=1.0\textwidth]{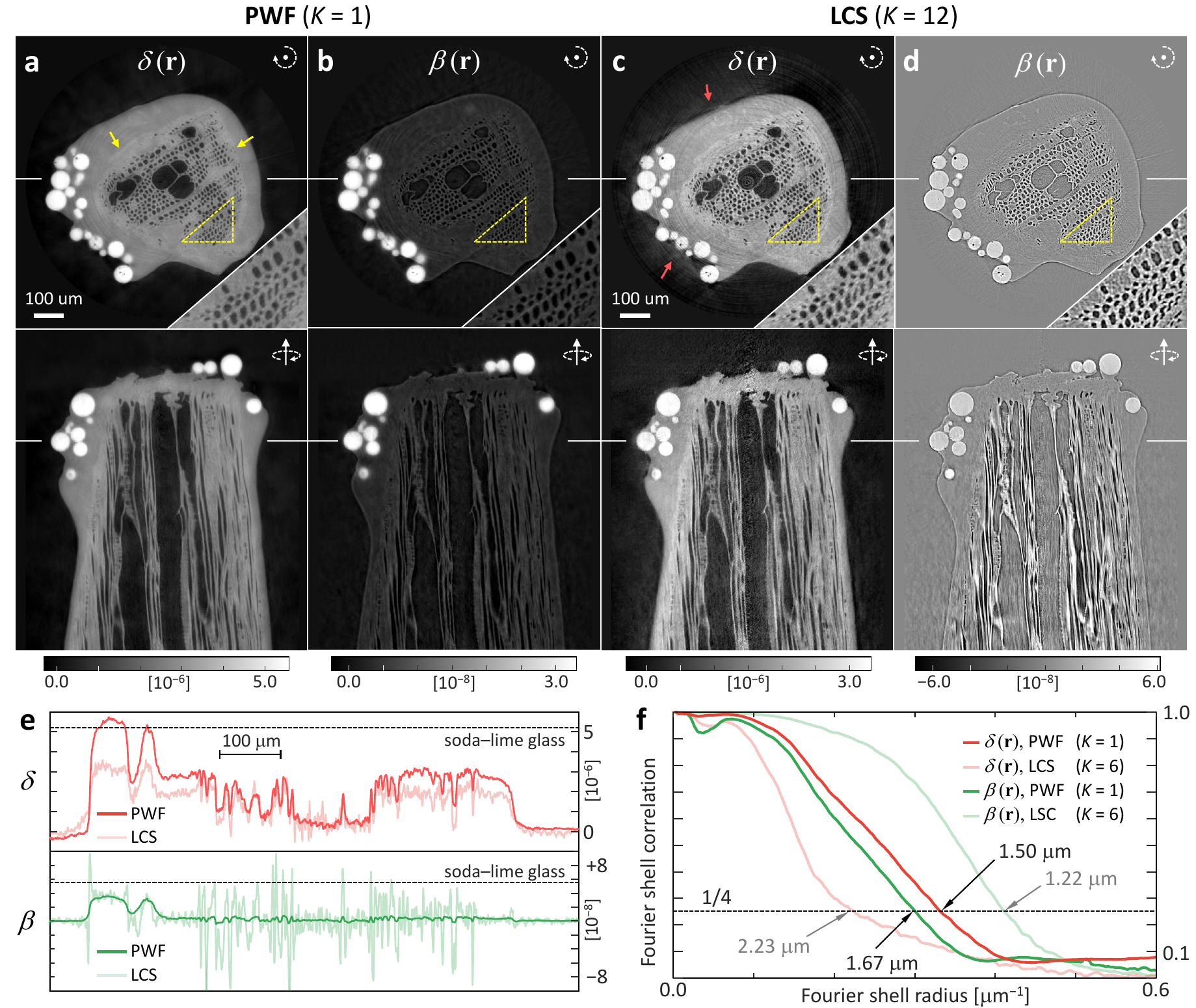}
\caption{
\textbf{Tomographic reconstruction results of a toothpick with glass beads.}
\textbf{a--b}, The real (a) and imaginary (b) parts of the reconstructed 3D refractive index distribution, $n(\mathbf{r}) = 1-\delta(\mathbf{r})+i\beta(\mathbf{r})$, from PWF with a single measurement ($K$ = 1).  Please refer to Video~\textcolor{blue}{S1} for the results from different cross sections. The yellow arrows indicate the interface between the toothpick and the adhesive. The white arrow symbols indicate the rotation axis orientations. The intersection of two orthogonal slices is shown as a solid line in the background.
\textbf{c--d}, The reconstructed counterparts from a LCS with 12 measurements in different diffuser positions ($K$ = 12) \cite{quenot2021implicit}. Note that the color scale is different from the PWF results (\textbf{a} and \textbf{b}). The red arrows indicate the reconstruction artifacts. 
\textbf{e}, Line profiles along the intersection lines shown in the backgrounds. The expected $\delta$ and $\beta$ values of the beads are shown as dotted lines. 
\textbf{f}, The Fourier shell correlation (FSC) of the reconstructed tomograms. The intersections with the 1/4 criterion (dotted line) are pointed by the arrows with the corresponding spatial resolutions.
}\label{fig3}
\end{figure}

From the field reconstruction results at various projection angles, we reconstruct the three-dimensional (3D) refractive index distribution of the sample, $n(\mathbf{r})=1-\delta(\mathbf{r})+i\beta(\mathbf{r})$, where $\mathbf{r}$ is 3D spatial position (see Methods). The $\delta(\mathbf{r})$ and $\beta(\mathbf{r})$ are separately reconstructed from the retrieved phase and attenuation images, respectively (Figs.~\ref{fig3}a--\ref{fig3}d). Please refer to Video~\textcolor{blue}{S1} for the results from different cross sections.

For both PWF and LCS, the $\delta(\mathbf{r})$ values are consistently higher than $\beta(\mathbf{r})$, resulting in better visualization of the detailed internal structure of the toothpick (Figs.~\ref{fig3}a and \ref{fig3}c). In PWF, the sample boundaries are clearly reconstructed, including the interface between the toothpick and the adhesive (Fig.\ref{fig3}a, yellow arrows). In LCS, although the sample structure is reconstructed to some extent, significant artifacts are present, suggesting inaccurate phase reconstruction (Fig.~\ref{fig3}c, red arrows). A clearer contrast between PWF and LCS can be seen in the $\beta(\mathbf{r})$ results. Unlike the clear distinction between the glass beads and the toothpick in PWF (Fig.~\ref{fig3}b), the LCS result shows large fluctuation along the edges (Fig.~\ref{fig3}d). Such fluctuation are mainly found at sample edges that produce large phase gradient (Fig.~\ref{fig2}g) and can be understood as phase-coupling artifacts.

We also compare the reconstructed $\delta$ and $\beta$ values of the soda–lime glass beads with those reported in the literature (Fig.~\ref{fig3}e). We calculate the expected $\delta$ and $\beta$ values from the density provided by the manufacturer \cite{henke1993x}. In PWF, we find that both $\delta$ and $\beta$ agree with the expected values, confirming the credibility of the reconstructed phase and attenuation values (Figs.~\ref{fig2}a and \ref{fig2}c). In LCS, however, significantly lower $\delta$ values are observed due to phase underestimation (Fig.~\ref{fig2}e). Highly fluctuating $\beta$ also provides physically incorrect negative $\beta$ values. These results also agree well with numerical simulation (Fig.~\ref{figS:PWFsimul_tomo}).

The 3D resolutions of the reconstructed tomograms are estimated based on Fourier shell correlation (FSC) analysis (Fig.~\ref{fig3}f, see Methods). We find that conventional resolution criterion (i.e., 1/7) is not directly applicable here due to nonlinear noise propagation in PWF algorithm. Instead, we use the 1/4 criterion based on our derivation (see Methods). Note that $K$ = 6 is used here for the LCS results in order to achieve two independent reconstructions from a total of 12 measurements (see Methods). Although FSC is useful for analyzing image resolution and comparing methods, it is a very sample-specific analysis. It does not present the general properties of an imaging system or algorithm. Therefore, direct extrapolation of the FSC results in Fig.~\ref{fig3}f to other samples or setups may be invalid.

For $\delta(\mathbf{r})$, we obtain a spatial resolution of \qtylist{1.50;2.23}{\um} for PWF and LCS, respectively. The significant improvement in spatial resolution strongly supports the direct reconstruction capability of the high-angle diffraction signal in PWF, which is considered as dark-field signals in LCS (Fig.~\ref{fig2}f). Although PWF reconstructs part of the dark field signal, we would like to clarify that not all scattering signals can be retrieved.

The theoretical spatial resolution of PWF is determined by the oversampling ratio criterion and the regularization window used. Since the regularization window is calculated based on the measured sample transfer function ($\mathrm{STF}_k$, see Methods), the achievable resolution is closely related to the speckle grain size. We experimentally demonstrate that the resolution of the reconstructed image degrades as the speckle grain size increases with different $L_2$ values of \qtylist{90;150}{\mm} (Fig.~\ref{figS:L2}). Despite the clear trend, establishing an universal theoretical limit on spatial resolution is difficult. Due to the smooth boundary of $\mathrm{STF}_k$ (Eq.~\ref{eq:STF}), reconstructed spatial bandwidth would highly depend on the practical noise level and the fineness of the sample structure. This is a common feature of projection-based X-ray imaging systems, which do not have band-limiting optical element between the sample and detector.

For $\beta(\mathbf{r})$, we have spatial resolutions of \qtylist{1.67;1.22}{\um} from PWF and LCS, respectively. In PWF, we achieve better spatial resolution in $\delta(\mathbf{r})$ than in $\beta(\mathbf{r})$, which encourages phase-contrast imaging not only for better contrast but also for better resolution. This also aligns well with the numerical simulation results (Fig.~\ref{figS:PWFsimul_tomo}). Interestingly, LCS shows better FCS results than PWF in $\beta(\mathbf{r})$. We observe the same behavior in the numerical simulation and determine that it is primarily caused by the strong, high-frequency phase-coupling artifacts of the LCS, which do not occur for the PWF (Fig.~\ref{figS:FSCsimul}). Further, the simulation results suggest that PWF has superior resolution in both $\delta(\mathbf{r})$ and $\beta(\mathbf{r})$ as shown in Figs.~\ref{figS:PWFsimul_tomo} and \ref{figS:FSCsimul}.

\begin{figure}[t]
\centering
\includegraphics[width=1.0\textwidth]{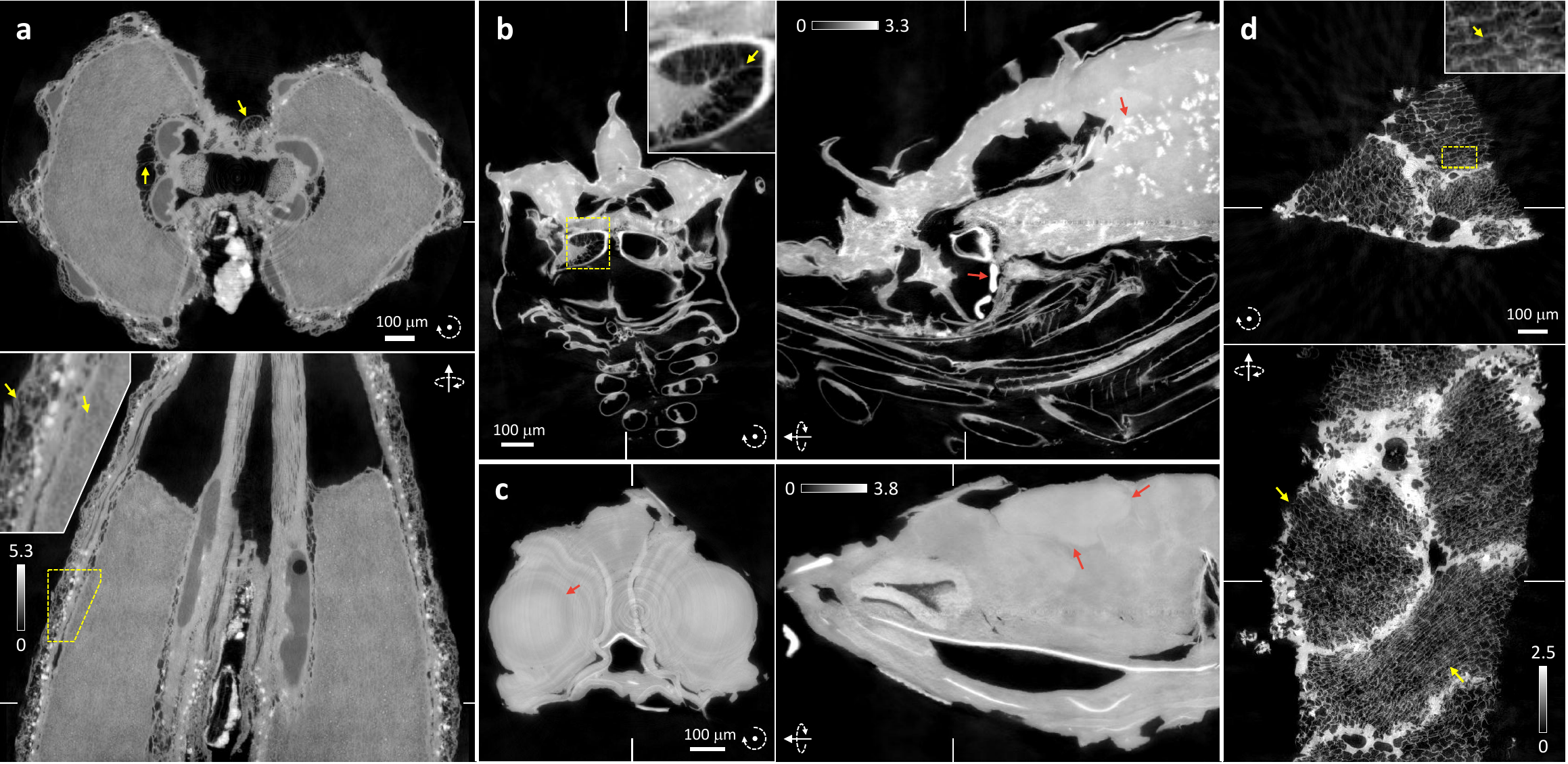}
\caption{
\textbf{Reconstructed quantitative phase tomogram $\delta(\mathbf{r})$ of various samples.} \textbf{a}, A cumin seed; \textbf{b}, a dried shrimp; \textbf{c}, a dried anchovy; and \textbf{d}, a piece of cork. The red and yellow arrows highlights the structural contrast of $\delta$, and fine structures, respectively. All colorbar units are $10^{-6}$. Please refer to Videos~\textcolor{blue}{S2--S5} for the results from different cross sections.  The reconstructed $\beta(\mathbf{r})$ are shown in Fig.~\ref{figS:fig4atten}.
}\label{fig4}
\end{figure}

To demonstrate the versatility of PWF, we also observe other samples---a cumin seed, a dried shrimp, a dried anchovy, and a piece of cork are chosen---with the same setup (Fig.~\ref{fig4}).
Due to the limited beamtime available, the additional samples are simply purchased from a local store and observed without any additional treatment (see Methods). All the experimental and reconstruction parameters are kept identical. 

We get surprisingly good reconstruction results for all the samples (Fig.~\ref{fig4}). Although we obtain results for both $\delta(\mathbf{r})$ and $\beta(\mathbf{r})$ for each sample, we focus more on $\delta(\mathbf{r})$ here, since $\beta(\mathbf{r})$ is achievable using traditional methods. It is noteworthy, however, significant differences between the $\delta(\mathbf{r})$ and $\beta(\mathbf{r})$ are found, which highlight the potential specificity enhancement of complex-field imaging (Fig.~\ref{figS:fig4atten}).

Fine sample structures are well visualized, as highlighted in the insets. Especially for the cumin seed, not only the outer fine structures but also the inner fine structures are clearly depicted (Fig.~\ref{fig4}a, yellow arrows). The fine and complex structures of the shrimp are also well reconstructed (Fig.\ref{fig4}b, yellow arrows). Lipids and the exoskeleton are naturally highlighted by the higher $\delta$ values (Fig.\ref{fig4}b, red arrows). We can also see the clear boundaries between the internal organs of the anchovy (Fig.~\ref{fig4}c, red arrows). The fine cell structures of the cork are successfully reconstructed (Fig.~\ref{fig4}d, yellow arrows). However, a certain amount of background artifacts are found in the cork result. We believe this is due to the mostly fine structure of cork, which effectively requires more projection angles for complete reconstruction \cite{kak2001principles}. Based on the Crowther criterion, at least 1047 projection angles in the range of \qtyrange{0}{180}{\degree} are required to fully reconstruct a \qty{1}{\mm} sample with a spatial resolution of \qty{1.5}{\um}, while we used 801 projection angles throughout the experiments.

\section{Discussion}\label{sec:Conclusion}
We successfully demonstrate speckle-based X-ray microtomography via PWF. To extract the sample field information from random speckle images, we combine three lessons learned from different disciplines: the Wirtinger flow is adapted from the random-measurement-based phase retrieval studies in signal processing and information theory \cite{candes2015phase, wang2017solving}; the preconditioner is inspired by the phase gradient retrieval of speckle-tracking methods studied in the field of X-ray phase contrast imaging \cite{morgan2012x, berujon2012two}; and the regularizer is inspired by the oversampling condition studied in the field of coherent imaging \cite{lee2016exploiting, levitan2020single}.

We present the detailed PWF algorithm, along with additional reconstruction results obtained without preconditioner or regularizer. In the experiments, we show that PWF provides good reconstruction results for both the phase and attenuation of the sample. The reconstruction fidelity and spatial resolution are confirmed by comparing the measured refractive index of a glass bead with the literature and FSC analysis. We also compare PWF with an existing speckle-tracking method (LCS) and find that PWF clearly outperforms LCS in terms of reconstruction fidelity and spatial resolution, despite the fact that all the comparison results are based on a single measurement for PWF, while 12 different measurements are used for LCS. Finally, we demonstrate the versatility of PWF by successfully visualizing 3D refractive index distributions of various samples without changing the setup and reconstruction parameters.

We believe that this work will facilitate speckle-based X-ray imaging in various applications. In synchrotron facilities, conventional microtomography beamlines can easily reproduce the same results by simply introducing a diffuser after the sample. We expect PWF to be applicable to a wide range of experimental conditions, regardless of source size and source-sample distance. We also expect that the spectral dimension can also be explored by scanning the X-ray energies without any changes to the imaging setup. The chemical or material composition can be inferred from the absorption and dispersion relations \cite{zhang2020depth}.

We are also optimistic about the application of PWF to benchtop X-ray sources, as the decoherence effects of partially coherent sources are already accounted for in the model. 
This extends the applicability of speckle-based phase-contrast X-ray imaging to a wider range of on-field applications. The single-shot nature of PWF also provides a significant practical advantage in terms of measuring time, stability, and radiation dose, especially in clinical applications \cite{bravin2012x}. 
We believe that the applicability to polychromatic sources should be explored first, since most benchtop sources provide polychromatic emission and the current forward model only considers monochromatic light.

\clearpage
\section{Materials and methods}\label{sec:Methods}

\subsection{Sample preparation}\label{subsec:samPrep}
Glass beads ($\leq\qty{106}{\um}$, $\qty{2.5}{\g\per\cm\tothe{3}}$, G8893, Merck KGaA) are attached to a toothpick with an adhesive for the sample shown in Figs.~\ref{fig2} and \ref{fig3}. The samples shown in Fig.~\ref{fig4} (a cumin seed, a dried anchovy, a dried shrimp, and a piece of cork) are procured from a local store. The piece of cork is obtained from a wine cork.

\subsection{Experimental setup}\label{subsec:expSetp}
The experiments are performed at the 6C beamline of PLS-II in Korea. A wiggler source is used (\qtyproduct{500 x 30}{\um} full width at half maximum), followed by a double multilayer monochromator (DMM) tuned to \qty{10}{\keV} ($\Delta E / E$ = 2.1\% measured at \qty{16}{\keV}). The sample is placed \qty{36}{\m} after the source. The diffuser is four layers of P3000 grit sandpaper (991A, Starcke GmbH \& Co. KG) placed $L_1$ = \qty{3}{\mm} after the sample. The scintillator is \num{50}-\unit{\um} thick LuAG:Ce placed $L_2$ = \qty{20}{\mm} downstream of the diffuser. Note that the shortest possible $L_1$ and $L_2$ are chosen to minimize sample-induced decoherence (Supplementary Text). Specifically, $L_1$ is chosen to ensure that it does not physically interfere with the sample rotation stage, and $L_2$ is determined to ensure that the spatial bandwidth of the speckle pattern is well sampled without aliasing. Note that the speckle grains become coarser as $L_2$ increases due to the partial coherence (Fig.~\ref{figS:speckle}). The image on the scintillator is observed with an optical microscope equipped with an objective lens ($\textrm{NA}$ = 0.25, LMPLFLN~10X, Olympus Corp.) and a sCMOS camera (\qty{6.5}{\um}, \numproduct{2560 x 2160}, pco.edge~5.5, Excelitas PCO GmbH). The corresponding spatial sampling period and area are \qty{650}{\nm} and \qtyproduct{1.664 x 1.404}{\mm}, respectively.

\subsection{Data acquisition}\label{subsec:dataAcq}
For each sample, data are acquired in the following order: (i) dark frame, (ii) flat field for the beam image, (iii) reference speckle from the diffuser, (iv) sample projection images. In step (iv), the samples are rotated continuously from \qtyrange{0}{180}{\degree}, while the projection images are taken on the fly at each \qty{0.225}{\degree} rotation angle, resulting in a total of 801 images. The camera exposure time is kept at \qty{100}{\ms} throughout the measurements. For speckle-tracking methods, which typically requires \numrange{10}{20} independent measurements \cite{zdora2020x,quenot2021implicit}, the acquisition process is repeated 12 times at different lateral positions of the diffuser. In all subsequent reconstruction steps, all speckle images are considered to have been pre-processed with dark-frame subtraction and flat-field correction. The entire acquisition process, including all stage movements and pauses, is completed in approximately \qty{4}{\min} (see Video~\textcolor{blue}{S6}).

\subsection{Intensity optical transfer function (IOTF)} \label{subsec:IOTF}
The coherence length at the detection plane is calibrated from the autocorrelation function of the reference speckle pattern based on the Siegert relation 
\begin{equation}
g^{(2)}(\Delta x) = 1 + \left| g^{(1)}(\Delta x) \right|^2, \label{sigert}
\end{equation}
while $g^{(1)}(\Delta x)$ and $g^{(2)}(\Delta x)$ are first- and second-order normalized autocorrelation functions of a speckle pattern \cite{goodman1979role}. Note the $\left| g^{(1)}(\Delta x) \right|^2$ term is closely related to the spatial coherence of light \cite{mandel1962measures}. Due to the inherent asymmetry of the synchrotron source, horizontal and vertical coherence lengths ($x_{\mathrm{coh}}$ and $y_{\mathrm{coh}}$) are estimated separately. For example, for $x_{\mathrm{coh}}$, we compute the one-dimensional normalized autocorrelation function along the $x$-axis, average it along the $y$-axis, and estimate the coherence length by fitting it to the Gaussian model $g^{(2)}(\Delta x) = 1+\exp\left(-\pi \Delta x^2/x_{\mathrm{coh}}^2\right)$ \cite{mandel1962measures}. Corresponding intensity point spread function (IPSF) becomes
\begin{equation}
\mathrm{IPSF}_r = \frac{2}{x_{\mathrm{coh}}y_{\mathrm{coh}}} \exp\left[-2\pi
\left( 
\frac{x^2}{x_{\mathrm{coh}}^2} + \frac{y^2}{y_{\mathrm{coh}}^2}
\right)
\right], \label{IPSF}
\end{equation}
where $r$ is the index of vectorized image in $(x,y)$ space. By the Fourier transform of Eq.~\ref{IPSF}, we can calculate the intensity optical transfer function (IOTF),
\begin{equation}
\mathrm{IOTF}_k = \exp\left[-\frac{\pi}{2}
\left( 
u^2 x_{\mathrm{coh}}^2 + v^2 y_{\mathrm{coh}}^2
\right)
\right]. \label{IOTF}
\end{equation}
where $u$ and $v$ are the spatial frequencies of $x$ and $y$, respectively, and $k$ is the index of vectorized image in $(u,v)$.

\subsection{Diffuser transmission function}\label{subsec:Diffuser}
We derived the transmission function of the diffuser ($t_r$) from the reference speckle. First, we removed the spatial incoherence effect using Wiener deconvolution with the calibrated IOTF (Eq.~\ref{IOTF}). This step estimates the `coherent' reference speckle intensity at the camera plane. Then, the $t_r$ can be obtained by the numerical backpropagation ($L_2$) of the deconvoluted reference speckle. One problem here is that its phase is unknown.

We originally attempted to retrieve the sample and diffuser fields simultaneously, as done in ptychographic iterations \cite{thibault2008high, maiden2009improved}. However, we found that the obtained diffuser phase is usually unreliable (e.g., slowly varying, spatially inhomogeneous) and changes drastically depending on the initial guess (Fig.~\ref{figS:diffPhase}). Despite this instability, the sample field was retrieved in a stable manner with indiscernible tomogram results. It is noteworthy that this result agrees with near-field ptychography results \cite{stockmar2013near, stockmar2015x, stockmar2015xtomo}. This may suggests that there is only one sample field that satisfies both the sample and reference speckle simultaneously, regardless of the phase of the reference speckle. In other words, the sample field can be retrieved with any of the possible diffusers phase. Based on this, we simply fix the phase of the reference speckle to zero. 

This is a significant advantage over conventional speckle-based imaging methods, which usually require a well-defined complex diffuser transmission function \cite{zhang2016phase, lee2023direct}. We expect this advantage to hold only for weakly scattering samples with a near-field setup that does not induce significant overlap between speckle grains. Further investigation is needed to determine these conditions.

\subsection{Preconditioning filter}\label{subsec:precondWin}
The phase retrieval problem in speckle-based X-ray microtomography is well conditioned on the phase gradient rather than the phase. As such, the gradient with respect to the phase gradient would have better convergence property. 
Thus, at each iteration, we want to perform the gradient descent step to the phase gradient such as
\begin{align} 
\frac{\partial \phi_r}{\partial x} \gets \frac{\partial \phi_r}{\partial x} 
- \eta\frac{\partial \mathcal{L}}{\partial (\partial \phi_r / \partial x)} \nonumber \\
\frac{\partial \phi_r}{\partial y} \gets \frac{\partial \phi_r}{\partial y} 
- \eta\frac{\partial \mathcal{L}}{\partial (\partial \phi_r / \partial y)}
\label{eq:iteration}
\end{align}
where $\mathcal{L}$ is the loss function defined in Eq.~\ref{eq:lossFunction}, $\phi_r=\mathrm{Im}(\psi_r)$ is the sample phase, and $\eta$ is a step size. Applying additional partial derivative to both sides of Eq.~\ref{eq:iteration} and adding them together, we get
\begin{equation} 
\nabla^2_\perp\phi_r \gets \nabla^2_\perp\phi_r - \eta \left(
 \frac{\partial}{\partial x} \frac{\partial \mathcal{L}}{\partial (\partial \phi_r / \partial x)}
+\frac{\partial}{\partial y} \frac{\partial \mathcal{L}}{\partial (\partial \phi_r / \partial y)}
\right), 
\label{eq:inverseGradient}
\end{equation}
where $\nabla^2_\perp=\partial^2/\partial x^2+\partial^2/\partial y^2$ is a two-dimensional Laplacian. The second term on the right hand size can be simplified by the chain rule
\begin{equation} 
\frac{\partial}{\partial x} \frac{\partial \mathcal{L}}{\partial (\partial \phi_r / \partial x)}
= \frac{\partial}{\partial x} \frac{\partial \phi_r}{\partial (\partial \phi_r / \partial x)} \frac{\partial \mathcal{L}}{\partial \phi_r}
= \frac{\partial (\partial \phi_r / \partial x)}{\partial (\partial \phi_r / \partial x)} \frac{\partial \mathcal{L}}{\partial \phi_r}
= \frac{\partial \mathcal{L}}{\partial \phi_r},
\end{equation}
which leads
\begin{equation} 
\nabla^2_\perp\phi_r \gets \nabla^2_\perp\phi_r - \eta \frac{\partial \mathcal{L}}{\partial \phi_r}, 
\label{eq:lapalacian}
\end{equation}
where the factor of 2 is omitted here as it is coupled with the step size. Applying inverse Laplacian to both sides of Eq.~\ref{eq:lapalacian}, we get
\begin{equation} 
\phi_r \gets \phi_r - \eta \nabla^{-2}_\perp\frac{\partial \mathcal{L}}{\partial \phi_r}, 
\label{eq:inverselapalacian}
\end{equation}
Compared to the conventional gradient-based phase iteration, $\phi_r \gets\phi_r-\eta\partial \mathcal{L}/\partial \phi_r$, there is an additional inverse Laplacian term before the gradient term, which is the preconditioner. The preconditioner is introduced by applying the inverse quadratic preconditioning filer $P^{-1}_k := 1/(u^2+v^2)$ to the update term,
\begin{equation} 
\phi_r \gets \phi_r - \mathcal{F}^{-1}
\left\{ 
    P^{-1}_k \mathcal{F}\left\{ \frac{\partial \mathcal{L}}{\partial \phi_r} \right\} 
\right\}
\label{eq:precondFilter},
\end{equation}
 where $u$ and $v$ are the spatial frequencies of $x$ and $y$, and $\mathcal{F}\left\{ \cdot \right\}$ is the Fourier transform. For the zero frequency ($u=0$ and $v=0$), $P^{-1}_k$ is set to zero, and is determined separately later by zeroing the background phase value. Since $\phi_r=\mathrm{Im}(\psi_r)$, the preconditioning filter is applied only to the imaginary part of $\psi_r$. The detailed calculation can be found in Algorithm~\ref{algorithm:functions}.

\subsection{Regularization window}\label{subsec:regWin}
In this work, the number of measured spatial modes is defined by the IOTF, which effectively limits the measurable spatial bandwidth. Since Eq.~\ref{IOTF} gives the effective bandwidths of $\sqrt{2}/x_{\mathrm{coh}}$ and $\sqrt{2}/y_{\mathrm{coh}}$ \cite{mandel1962measures}, we can estimate the number of measured spatial modes from the space-bandwidth product (SBP) of the measured images \cite{lee2023direct}, 
\begin{equation}
M 
= \frac{1}{2}
\left(\frac{\sqrt{2}\mathrm{FOV}_x}{x_{\mathrm{coh}}}\right)
\left(\frac{\sqrt{2}\mathrm{FOV}_y}{y_{\mathrm{coh}}}\right)
= 
\left(\frac{\mathrm{FOV}_x}{x_{\mathrm{coh}}}\right)
\left(\frac{\mathrm{FOV}_y}{y_{\mathrm{coh}}}\right)
, \label{eq:M}
\end{equation}
where $\mathrm{FOV}_x$ and $\mathrm{FOV}_y$ are the horizontal and vertical extents of acquired images (i.e., field of view). Note that the factor $1/2$ is introduced here to correctly estimate the SBP of speckle field from the SBP of intensity speckle image. Assuming the Gaussian angular spectrum, the intensity speckle would have greater bandwidth by the factor of $\sqrt{2}$, which results in the SBP increase by the factor of 2. 
The number of reconstructed spatial modes ($N$) can be defined from given oversampling ratio $\gamma$ and Eq.~\ref{eq:M},
\begin{equation}
N = 
\left(\frac{\mathrm{FOV}_x}{ \sqrt{\gamma} x_{\mathrm{coh}}}\right)
\left(\frac{\mathrm{FOV}_y}{ \sqrt{\gamma} y_{\mathrm{coh}}}\right). \label{eq:N}
\end{equation}
Since the reconstructed fields have the same spatial extent as the measured images, Eq.~\ref{eq:N} directly leads to additional constraints on their Fourier space by the bandwidth of $x^{-1}_{\mathrm{coh}}/\sqrt{\gamma} $ 
and $y^{-1}_{\mathrm{coh}}/\sqrt{\gamma} $. Assuming a Gaussian profile, we can define the sample transfer function ($\mathrm{STF}_k$),
\begin{equation}
\mathrm{STF}_k= 
\exp\left[-\pi\gamma 
\left( u^2 x_{\mathrm{coh}}^2 + v^2 y_{\mathrm{coh}}^2 \right)
\right]  \label{eq:STF}
\end{equation}
and subsequently the regularization window, 
\begin{equation}
    \Gamma_k^2 = 1 - \mathrm{STF}_k. \label{eq:regWin}
\end{equation}
The used oversampling ratio ($\gamma = 1$) may seem inconsistent with previous works that empirically report $\gamma \geq 4$ for robust field reconstruction \cite{candes2015phase, lee2016exploiting}. The difference is that here we use a Gaussian STF, which provides a soft boundary, while previous works used a well-defined domain, which is equivalent to binary STFs. We find that the soft boundary relaxes the dependence of the reconstruction fidelity on $\gamma$ (Fig.~\ref{figS:gamma}), and we prefer to use a smaller $\gamma$ to obtain the best possible resolution.

Due to the Gaussian shape of the $\mathrm{STF}_k$, it does not have a distinct limit on the resolution; rather, it depends heavily on the fineness of the sample structure and the noise level. For instance, if the sample bandwidth is retrievable up to $\mathrm{STF}_k = 0.1$ boundary, the expected spatial resolution can be calculated as $\sqrt{\pi}x_\mathrm{coh}/\left(2\log10\right) \approx 0.385 x_\mathrm{coh}$ from Eq.~\ref{eq:STF}. Using the measured coherence lengths in Fig.~\ref{figS:speckle}, the expected horizontal and vertical resolution becomes \qtylist{1.34;1.66}{\um}, respectively. The root-mean-square resolution is \qty{1.51}{\um}, which agrees with the acquired resolution in FSC analysis (Fig.~\ref{fig3}f).

\subsection{Reconstruction procedure}\label{subsec:tomorecon}
The PWF iteration is stopped when the normalized correlation between the retrieved field of the current iteration and the previous iteration is greater than $10^{-0.00001}$. Typically, \numrange{700}{1300} iterations are required to meet the criterion, which takes \qtyrange{5}{6}{\s} on a personal computer equipped with a graphics processing unit (GPU; GeForce RTX 4090, NVIDIA Corp.) using MATLAB software (The MathWorks, Inc.). Thus, for each sample, it took \qtyrange{70}{80}{\min} to process all 801 projection angles. Note that all the computations in Algorithm~\ref{algorithm:PWF} are element-wise operations and Fourier transforms, which were effectively accelerated by the GPU.

In both PWF and LCS, the root-mean-square error (RMSE) image is calculated from the reconstructed sample fields. By applying reconstructed sample fields to the forward model, we calculate the estimated speckle patterns, $f(x_r)$ and subtract it from the measured speckle patterns, $y_r$. In PWF, there is only one speckle pattern to compare ($K=1$), so the RMSE image simply is $|y_r-f(x_r)|$. In LCS, we have twelve speckle patterns to compare ($K=12$), so the RMSE image becomes $\sqrt{\sum_{m=1}^K\left(y^m_r-f_\mathrm{LCS}(x^m_r)\right)^2}$, where $f_\mathrm{LCS}$ is the forward model of LCS.
The RMSE images are normalized with the mean intensity of the reference speckle and present in \% (Figs.~\ref{fig2}d and \ref{fig2}h). 

The tomogram results are reconstructed using the standard filtered back projection (FBP) algorithm with Ramachandran-Lakshminarayanan (``Ram–Lak'') filter. The \textit{iradon} function of MATLAB is independently applied at each vertical position of the sample. A constant $\lambda/p/(2\pi)$ is then multiplied to convert the tomogram results to $\delta(\mathbf{r})$ and $\beta(\mathbf{r})$, where $p$ is a pixel size of the reconstructed sample tomogram. We often experience low-frequency background curvature due to the intrinsic lower sensitivity on slow phase variation of phase-gradient sensing techniques \cite{choi2017compensation}. We mitigate such artifacts by fitting the background area with a two-dimensional quadratic function after tomographic reconstruction. 

\subsection{Fourier shell correlation (FSC)}\label{subsec:FRC}
For FSC, we utilize two tomograms of the toothpick sample (Fig.~\ref{fig3}) that are reconstructed from the measurements at different lateral positions of the diffuser. Defining either the real or imaginary parts of two tomograms as $g_1(\mathbf{r})$ and $g_2(\mathbf{r})$, the FSC can be calculated from the normalized cross-correlation of their Fourier transforms over the shell,
\begin{equation}
\mathrm{FSC}(\kappa) = 
\frac
{ \left\langle  \tilde{g}_1^*(\mathbf{k})\tilde{g}_2(\mathbf{k}) \right\rangle_{\|\mathbf{k}\|_2=\kappa}  }
{ \sqrt{ 
\left\langle 
    \tilde{g}_1^*(\mathbf{k})
    \tilde{g}_1(\mathbf{k})  
\right\rangle_{\|\mathbf{k}\|_2=\kappa} 
\left\langle  
    \tilde{g}_2^*(\mathbf{k})
    \tilde{g}_2(\mathbf{k}) 
\right\rangle_{\|\mathbf{k}\|_2=\kappa} 
} }, \label{eq:FSC}
\end{equation} 
where $\tilde{g}_{1,2}(\mathbf{k})$ is the Fourier transform of $g_{1,2}(\mathbf{r})$, and $\kappa$ is the radius of the shell in the 3D Fourier space \cite{rosenthal2003optimal}. For PWF, a total of 12 tomograms are independently reconstructed from the $K$ = 12 measurements, and two are randomly selected for FSC calculation. Five FSC are calculated from different pairs of tomograms and averaged. For LCS, the $K$ = 12 measurements are randomly divided into two sets of $K$ = 6, and the tomograms reconstructed from the two sets are used for the FSC. Similarly, five FSC are calculated from different random splits and averaged.

Regarding the resolution criterion, we followed the derivation in Ref.~\cite{rosenthal2003optimal}, which suggests the 0.143 (or 1/7) criterion for the ``full-set averaged'' result. For example, if there are $K$ repetitive measurements, the FSC is calculated between the two half-set averaged results of $K/2$ non-overlapping measurements. To extrapolate this FSC result to the resolution of the full-set averaged result, an improvement in signal-to-noise ratio (SNR) by $\sqrt{2}$ were assumed. Unfortunately, this result cannot be applied directly to our case for two reasons: first, PWF requires only a single measurement ($K$ = 1), which cannot be divided into two sets; and more importantly, the SNR may not be simply proportional to $\sqrt{K}$ because noise propagation in PWF is more complex due to the nonlinear reconstruction process. Therefore, we redo the derivation without the full set extrapolation.

Let us divide the measured tomogram into the sample signal and noise terms, $g_{1,2} (\mathbf{r})=s_r(\mathbf{r})+n_{1,2}(\mathbf{r})$. Note that $s_r(\mathbf{r})$ is invariant across measurements, while $n_{1,2}(\mathbf{r})$ fluctuates randomly. Here we can consider the result to be reliable if the normalized correlation between the data and the ideal sample signal is greater than $1/2$,
\begin{equation}
\frac{1}{2} \leqq \frac
{ \left\langle 
\tilde{g}^\ast\tilde{s}
\right\rangle  }
{ 
    \sqrt{
        \left\langle \tilde{g}^\ast\tilde{g}  \right\rangle 
        \left\langle \tilde{s}^\ast\tilde{s}  \right\rangle
    }
}
 =
 \sqrt{
    \frac{
        \left\langle 
        \tilde{s}^\ast\tilde{s}
        \right\rangle  
    }
    {
        \left\langle \tilde{s}^\ast\tilde{s}  \right\rangle + 
        \left\langle \tilde{n}^\ast\tilde{n}  \right\rangle
    }
},
\label{eq:C=0.5}
\end{equation}
where the ${\|\mathbf{k}\|_2=\kappa}$ condition of the average operator, and the argument of $\tilde{g}(\mathbf{k})$, $\tilde{s}(\mathbf{k})$, and $\tilde{n} (\mathbf{k})$ are dropped for brevity. Note that the $\left\langle \tilde{n}^\ast\tilde{s} \right\rangle$ and $\left\langle \tilde{s}^\ast\tilde{n} \right\rangle$ terms have disappeared since the noise should be random and uniform over space, which leads
$\left\langle \tilde{s}^\ast\tilde{n} \right\rangle
=
\left\langle \tilde{s}^\ast \right\rangle
\left\langle \tilde{n} \right\rangle $ and $\left\langle \tilde{n} \right\rangle=0$ for the non-zero spatial frequencies ($\kappa >0$), respectively. Similarly, we can rewrite Eq.~\ref{eq:FSC} as
\begin{equation}
\mathrm{FSC}(\kappa) = \frac
{ \left\langle 
\tilde{s}^\ast\tilde{s}
\right\rangle  }
{ 
\left\langle \tilde{s}^\ast\tilde{s}  \right\rangle + 
\left\langle \tilde{n}^\ast\tilde{n}  \right\rangle
 }. \label{eq:FSC+SN}
\end{equation}
Substituting Eq.~\ref{eq:FSC+SN} into Eq.~\ref{eq:C=0.5}, we get $\mathrm{FSC}(\kappa)\geqq 1/4$, which suggests $1/4$ as the resolution criterion. Again, the only difference from the derivation in Ref.~\cite{rosenthal2003optimal} is the full set extrapolation that introduces additional $1/2$ factor for $\left\langle \tilde{n}^\ast\tilde{n}  \right\rangle$ in Eq.~\ref{eq:C=0.5}.




\subsection*{Data availability}
The MATLAB codes and experimental data shown in Fig.~\ref{fig2} are available here: https://doi.org/10.5281/zenodo.15073834. Full data are available on request from the corresponding authors.

\subsection*{Acknowledgements}
 This work was supported by National Research Foundation of Korea grant funded by the Korea government (MSIT) (RS-2024-00442348, 2021R1C1C2009220, 2022M3H4A1A02074314), the Ministry of Trade, Industry and Energy (MOTIE) and Korea Institute for Advancement of Technology (KIAT) through the International Cooperative R\&D program (P0028463), and the Korean Fund for Regenerative Medicine (KFRM) grant funded by the Korea government (the Ministry of Science and ICT and the Ministry of Health \& Welfare) (21A0101L1-12). Experiments using PLS-II were supported in part by MSIT and POSTECH. The authors gratefully acknowledge the support of Mr. Seob-Gu Kim and Dr. Yong Sung Park during the beamline experiments.

\subsection*{Conflict of interest}
The authors declare no conflicts of interest

\subsection*{Contributions}
K.L. proposed the idea, developed the principles and algorithm, designed experimental setup, conducted experiments, analyzed the data, and contributes to paper writing; 
H.H. developed the principles and algorithm, designed experimental setup, conducted experiments, analyzed the data, and contributes to paper writing; 
J.L. conducted experiments;
Y.P. supervised the project and contributes to paper writing.

\bibliography{bib_Xray6C}

\newpage
\renewcommand{\thefigure}{S\arabic{figure}}
\renewcommand{\thetable}{S\arabic{table}}
\renewcommand{\theequation}{S\arabic{equation}}
\renewcommand{\thealgorithm}{S\arabic{algorithm}}
\renewcommand{\thepage}{S\arabic{page}}

\renewcommand{\theHfigure}{S\arabic{figure}}
\renewcommand{\theHtable}{S\arabic{table}}
\renewcommand{\theHequation}{S\arabic{equation}}
\renewcommand{\theHalgorithm}{S\arabic{algorithm}}

\setcounter{figure}{0}
\setcounter{table}{0}
\setcounter{equation}{0}
\setcounter{algorithm}{0}
\setcounter{page}{1} 

\begin{center}
\section*{Supplementary Information for:\\ \paperTitle}

KyeoReh~Lee$^{\ast\dagger}$,
Herve~Hugonnet$^\dagger$,
Jae‑Hong~Lim,
and YongKeun~Park$^\ast$\\

\small$^\ast$Corresponding authors. Email: lee.kyeo@gmail.com; yk.park@kaist.ac.kr\\
\small$^\dagger$These authors contributed equally to this work.
\end{center}

\subsubsection*{This PDF file includes:}
Supplementary Text\\
Table S1\\
Algorithm S1 to S2\\
Figs. S1 to S12\\
Captions for Videos S1 to S6\\

\newpage
\subsection*{Supplementary Text}
\subsubsection*{Theory on coherent speckle generation}
\begin{figure}[ht]
\centering
\includegraphics[width=0.6\textwidth]{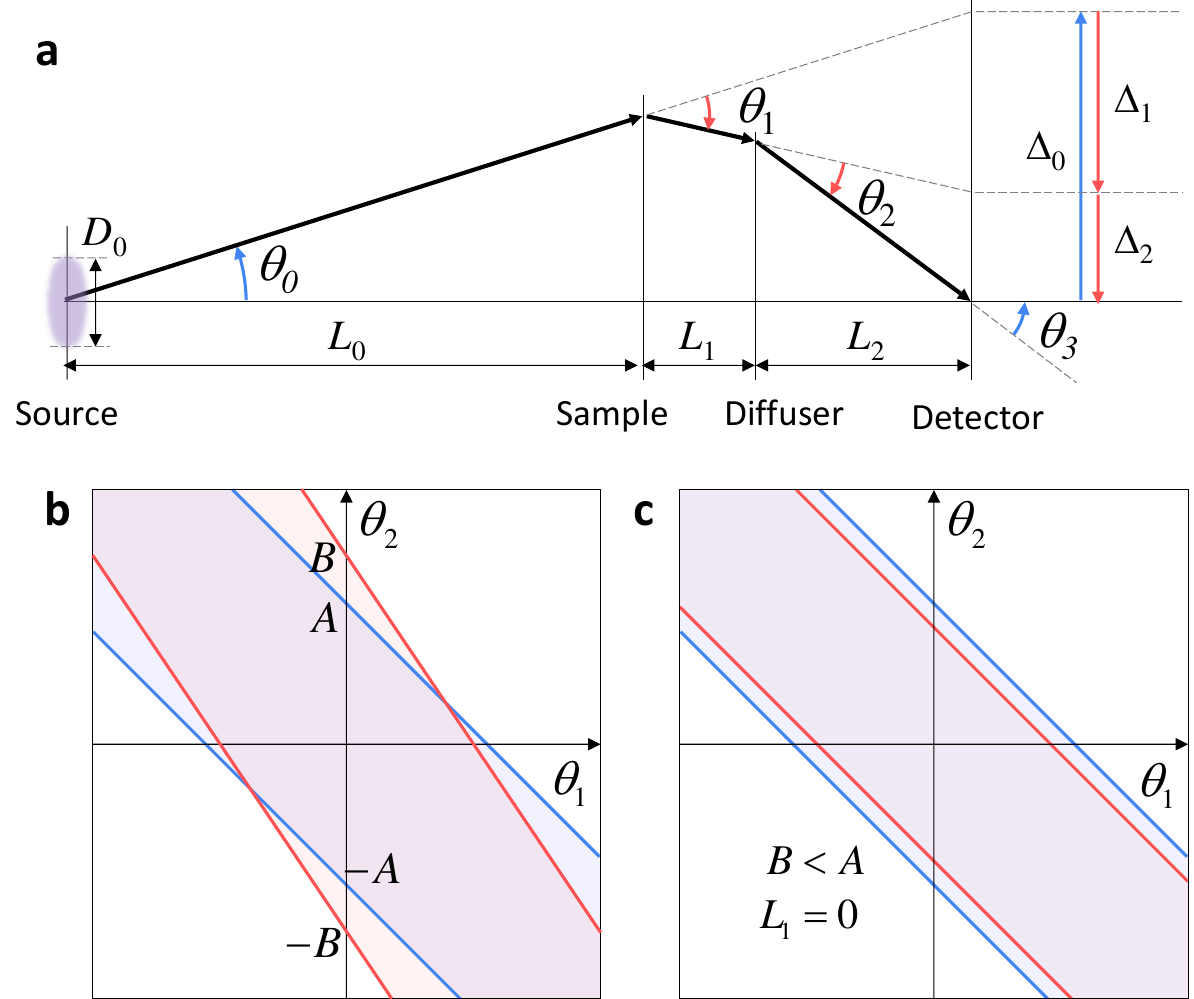}
\caption{
\textbf{Coherence speckle generation in partially coherent X-ray microtomography}
\textbf{a}, Setup diagram. The source, sample, diffuser and detector are placed from the front. The definitions of the variables used are as follows: $D_0$, the size of the source; $\theta_0$, the emission angle of the source; $\theta_1$, the diffraction angle of the sample; $\theta_2$, the diffraction angle of the diffuser; $\theta_3$, the detection angle at the detector; $\Delta_0$, $\Delta_1$, and $\Delta_2$ are the lateral propagation distances due to $\theta_0$, $\theta_1$, and $\theta_2$, respectively; and $L_0$, $L_1$, and $L_2$ are the distances between the elements. Note that the angles can be negative. The blue and red arrows represent positive and negative angles, respectively.
\textbf{b}, Valid $(\theta_1,\theta_2)$ pairs to construct a speckle grain. The blue and red lines represent the bounds of the equations ~\ref{eqS:thetaConditionPixel} and \ref{eqS:thetaConditionSource}, respectively, and the shaded area represents the $(\theta_1,\theta_2)$ pairs that satisfy the equations. The constants $A$ and $B$ are the intercepts of $\theta_2$, which are $\lambda/(4p) + \lambda/(2D_0)$ and $\lambda(L_0+L_1)/(2D_0L_2) + \lambda/(2D_0)$, respectively.
\textbf{c}, Same plot as \textbf{b}, but when $B<A$ and $L_1=0$, the overlapped area is constrained only by Eq.~\ref{eqS:thetaConditionSource}.
}\label{figS:theory}
\end{figure}

Consider a general speckle-based X-ray microtomography setup consisting of an X-ray source, a sample, a diffuser, and a detector (Fig.~\ref{figS:theory}).
We draw a straight line from the source to any point on the detector and describe the diffractive angles from the sample and the diffuser relative to this line, $\theta_1$ and $\theta_2$, respectively (Fig.~\ref{figS:theory}a). The corresponding emission ($\theta_0$) and detection ($\theta_3$) angles are determined based on two geometric relations
\begin{equation}
    \theta_0+\theta_1+\theta_2+\theta_3=0 \label{eqS:geoRelAngles}
\end{equation}
and
\begin{equation}
    \Delta_0+\Delta_1+\Delta_2=0, \label{eqS:geoRelDistances}
\end{equation}
where $\Delta_0$, $\Delta_1$, and $\Delta_2$ are the lateral propagation distances due to $\theta_0$, $\theta_1$, and $\theta_2$, respectively (Fig.~\ref{figS:theory}a). Under the paraxial approximation, we have $\Delta_0 = \theta_0 (L_0 + L_1 + L_2)$, $\Delta_1 = \theta_1 (L_1 + L_2)$, and $\Delta_2 = \theta_2 L_2$, where $L_0$, $L_1$, and $L_2$ are the source-sample, sample-diffuser, and diffuser-detector distances, respectively (Fig.~\ref{figS:theory}a).

To produce the coherent speckle pattern at the detector plane, the detector should not be able to distinguish our source from a coherent one. To achieve this, the emission angles should be small enough that the corresponding diffraction limit is larger than the source size ($D_0$)
\begin{equation} 
    \frac{\lambda}{2\left| \theta_0 \right|} > D_0, \label{eqS:coherenceCond} 
\end{equation} 
where $\lambda$ is the X-ray wavelength. To capture the speckle pattern without loss, the detection angle should be smaller than the Nyquist frequency of the detector ($0.5/p$), where $p$ is the effective pixel size of the detector. Since we are measuring the intensity part of the speckle, which doubles the bandwidth, we have 
\begin{equation} 
    \frac{2\left| \theta_3 \right|}{\lambda} < \frac{1}{2p}. \label{eqS:sampleCond} 
\end{equation}
Here, we do not consider the sizes of the sample and diffuser, assuming they are much larger than $\theta_0L_0$ and $\theta_3L_2$, respectively. If the source is coherent enough, the size of the sample or diffuser matters, as discussed in Ref.~\cite{lee2023direct}.

Substituting Eqs.~\ref{eqS:coherenceCond} and \ref{eqS:sampleCond} into Eqs.~\ref{eqS:geoRelAngles} and \ref{eqS:geoRelDistances}, we can derive two conditions on $\theta_1$, and $\theta_2$,
\begin{equation} 
    \left| \theta_1 + \theta_2 \right| < A
    \label{eqS:thetaConditionPixel} 
\end{equation}
and
\begin{equation} 
    \left|
        \left(1+\frac{L_1}{L_2}\right)\theta_1 + \theta_2 
    \right| < B, 
    \label{eqS:thetaConditionSource} 
\end{equation}
where $A = \lambda/(4p) + \lambda/(2D_0)$ and $B=\lambda(L_0+L_1)/(2D_0L_2)+ \lambda/(2D_0)$ are the $\theta_2$-intersects.

Two conditions are presented in Fig.~\ref{figS:theory}b as blue and red shaded areas with corresponding boundaries of the same colors (solid lines). The overlapped area represents the valid $(\theta_1, \theta_2)$ pairs that can construct coherent speckle on the detector. Note that if $\theta_1$ and $\theta_2$ have different signs, they can cancel each other and generate an acquirable coherent speckle. This is analogous to the analyzer grating in grating shearing interferometry \cite{weitkamp2005x}. Since the red lines always have a steeper slope than the blue lines, the two lines always intersect, effectively reducing the overlapped area. In other words, $L_1 = 0$ is preferred to maximize the overlapped area. This is why we keep $L_1$ at a minimum throughout the experiments.

Since only Eq.~\ref{eqS:thetaConditionPixel} condition contains the pixel size, violating this condition would induce finer speckles than the pixel size, which is not preferred for computational processing due to aliasing. Therefore, we find it better to be constrained by Eq.~\ref{eqS:thetaConditionSource} rather than by Eq.~\ref{eqS:thetaConditionPixel} as depicted in Fig.~\ref{figS:theory}c. To achieve that condition, we need to make $B < A$ with $L_1=0$, which results in 
\begin{equation}   
    2p\frac{L_0}{D_0} < L_2, 
    \label{eqS:L2condition} 
\end{equation}
which defines minimum $L_2$. In our experimental situation ($p=\qty{650}{\nm}$, $D_0=\qty{500}{\um}$, and $L_0=\qty{36}{\m}$) the minimum $L_2$ is \qty{93.6}{\mm}, 
which is inconsistent with the measured speckle patterns (Fig.~\ref{figS:speckle}). We believe this discrepancy arises because the experimental $L_0/D_0$ differs from the theoretical value due to experimental perturbations and diffraction from the downstream optics, a known problem with the beamline we used.

\clearpage
\renewcommand*\thefootnote{\alph{footnote}}
\newcolumntype{L}[1]{>{\hsize=#1\hsize\raggedright\arraybackslash}X}%
\newcolumntype{R}[1]{>{\hsize=#1\hsize\raggedleft\arraybackslash}X}%
\newcolumntype{C}[1]{>{\hsize=#1\hsize\centering\arraybackslash}X}%

\begin{table}[h]
\caption{Speckle-based phase retrieval methods}\label{tab:STMcomparative}
\begin{tabularx}{\textwidth}{@{\extracolsep\fill}L{0.3}C{0.8}L{1.2}L{0.5}L{0.7}L{2.5}}
\toprule%
Method & Refs. & Assumption& Min. $K$ & Computing time\footnotemark[1] & Remarks \\
\midrule\midrule
UMPA  & \cite{zanette2014speckle, zdora2017x, zdora2020x, savatovic2023multi}  & - & 1\footnotemark[2] & \qty{15.2}{\min}\footnotemark[3]& Slow reconstruction speed. Image resolution and quality highly depend on the window size \\    \midrule
GF    & \cite{paganin2018single}  & Phase only  & 1  & \qty{0.1}{\s}\footnotemark[4] & The assumption results in no attenuation image. \\
\midrule
MIST  & \cite{pavlov2020x,alloo2023m} & Single material \cite{alloo2023m} & 4\footnotemark[2] 
&\qty{8.0}{\s}\footnotemark[3]$^,$\footnotemark[5] & The assumption results in coupled attenuation and phase images.\\ 
\midrule
LCS  & \cite{quenot2021implicit, magnin2023dark} & - & 3\footnotemark[2] & \qty{0.2}{\s}\footnotemark[3] & Strong phase-coupling artifacts in the attenuation image. \\ 
\midrule
PWF  & This work  & - & 1  &\qty{4.5}{\s}\footnotemark[4] & Iterative algorithm \\
\botrule
\end{tabularx}
\footnotetext[a]{For single precision images (\qtyproduct{1630 x 1487}{}) using MATLAB software with a GPU (GeForce RTX 4090, NVIDIA Corp.), unless specified otherwise.}
\footnotetext[b]{$K>10$ is used in most practical demonstrations for acceptable signal-to-noise level.}
\footnotetext[c]{$K=12$}
\footnotetext[d]{$K=1$}
\footnotetext[e]{The GPU is not used here since the original Python script from Ref.~\cite{alloo2023m} is used as is.}
\end{table}

\clearpage
\begin{algorithm} [ht]
\caption{ Preconditioned Wirtinger flow (PWF). The vector indices $r$ and $k$ are used for real and reciprocal spaces, respectively. Please refer to Algorithm~\ref{algorithm:functions} for detailed steps in the functions used.
} \label{algorithm:PWF}
\begin{algorithmic}[1]
\State \textbf{Input:} 
    Measured sample speckle $\{y_r\}\in\mathbb{R}^m$; diffuser transmission function $\{t_r\}\in\mathbb{C}^m$; X-ray wavelength $\lambda$; propagation lengths $L_1, L_2\in\mathbb{R}$; intensity optical transfer function $\left\{\mathrm{IOTF}_k\right\}\in\mathbb{R}^m$; preconditioning filter $\{P^{-1}_k\}\in\mathbb{R}^m$; regularization window $\{\Gamma^2_k\}\in\mathbb{R}^m$; complex regularization parameter $\alpha\in\mathbb{C}$; step size $\eta\in\mathbb{R}$; and $m\in\mathbb{R}$ is the number of image pixels.
\Statex

\Procedure {PWF}{$ y_r, t_r,\lambda,L_1, L_2, \mathrm{IOTF}_k, P^{-1}_k, \Gamma^2_k, \alpha, \eta $} 
    \State $\psi_r \gets 0$ \Comment{ Complex phase, $\psi_r = \log x_r$ }
    \State $\phi_r \gets 0$ 
    \State $\mu \gets 1$ 

    \While{$\psi_r$ not converged}   
        \State $\phi^\mathrm{prev}_r \gets \phi_r$        
        \State $\mu^\mathrm{prev} \gets \mu$
    
        \State $f(x_r), x_r, v_r \gets$ \Call{PhysicalModel}{$\psi_r, t_r,\lambda, L_1, L_2, \mathrm{IOTF}_k$}
        \State $e_r \gets y_r - f(x_r)$           
        \State $g_r \gets$ \Call{WirtingerDerivative}{$e_r, x_r, v_r, t_r, \lambda, L_1, L_2, \mathrm{IOTF}_k$}

        \State $g_r \gets$ \Call{Preconditioner}{$g_r, P^{-1}_k$}
        $+$ \Call{Regularizer}{$\psi_r, \Gamma^2_k, \alpha$}

        \State $\phi_r \gets \psi_r-\eta g_r$
        \State $\mu \gets \frac{1+\sqrt{1+4\mu^2}}{2}$
        \State $\psi_r \gets \phi_r 
        + \left( \frac{\mu^\mathrm{prev}-1}{\mu}   \right)
          \left( \phi_r-\phi^\mathrm{prev}_{\mathbf{r}} \right)$
          \Comment{Nesterov's accelerated gradient \cite{nesterov1983method}}
        
    \EndWhile    
    \State \textbf{return} $\psi_r$
\EndProcedure
\end{algorithmic}
\end{algorithm}

\begin{algorithm} [ht]
\caption{ Used functions in Algorithm~\ref{algorithm:PWF}. } \label{algorithm:functions}
\begin{algorithmic}[1]
\Function {PhysicalModel}{$\psi_r, t_r, \lambda, L_1, L_2, \mathrm{IOTF}_k$}
    \State $x_r \gets e^{\psi_r}$
    \State $v_r \gets $ \Call{FreePropagation}{$x_r, \lambda,L_1$}
    \Comment{Sample to diffuser} 
    \State $v_r \gets $ \Call{FreePropagation}{$t_r v_r, \lambda,L_2$}  
    \Comment{ Diffuser to detector } 
    \State $f(x_r) \gets$ \Call{Filter}{$\left| v_r \right|^2, \mathrm{IOTF}_k$}    
    \Comment{ IPSF convolution }
    \State \Return $f(x_r), x_r, v_r$
\EndFunction
\Statex

\Function {WirtingerDerivative}{$e_r, x_r, v_r, t_r, \lambda, L_1, L_2, \mathrm{IOTF}_k$}
    \State $g_r \gets$ \Call{Filter}{$-e_r, \mathrm{IOTF}_k$}
    \State $g_r \gets$ \Call{FreePropagation}{$v_rg_r,\lambda,-L_2$} 
    \State $g_r \gets$ \Call{FreePropagation}{$t^*_rg_r,\lambda, -L_1$} 
    \State $g_r \gets x^*_rg_r$

    \State $g_r \gets g_r/\max_r |t_r|^2$ 
    \State \Return $g_r$
\EndFunction
\Statex

\Function {Preconditioner}{$g_r, P^{-1}_k$}
    \State $g''_r \gets \mathrm{Im}\left(g_r\right)$ 
    \Comment{Gradient for the sample phase} 
    \State $g''_r \gets $ \Call{Filter}{$g''_r, P^{-1}_k$}
    \State $g_r \gets \mathrm{Re}\left(g_r\right)+ig''_r$ 
    \State \Return $g_r$
\EndFunction
\Statex

\Function {Regularizer}{$\psi_r, \Gamma^2_k, \alpha$}
    \State $\rho_r \gets $ \Call{Filter}{$\psi_r, \Gamma^2_k$}
    \State $\rho_r \gets 
    \mathrm{Re}(\alpha)
    \mathrm{Re}(\rho_r)
    +i
    \mathrm{Im}(\alpha)
    \mathrm{Im}(\rho_r)$ 
    \State \Return $\rho_r$
\EndFunction
\Statex

\Function {FreePropagation}{$x_r, \lambda, L$}
    \State $w \gets \sqrt{\lambda^{-2}-u^2-v^2}$
    \Comment{Spatial frequency along the propagation direction} 
    \State $Q_k \gets e^{i2\pi wL}$ 
    \State $x_r \gets $ \Call{Filter}{$x_r, Q_k$}  
    \State \Return $x_r$
\EndFunction
\Statex

\Function {Filter}{$x_r, W_k$}
    \State $x_r \gets 
    \mathcal{F}^{-1}\left\{
    \mathcal{F}\left\{ x_r \right\} W_k 
    \right\} $  
    \Comment{$\mathcal{F}\{\cdot\}$ denotes the Fourier transform} 

    \State \Return $x_r$
\EndFunction
\end{algorithmic}
\end{algorithm}

\begin{figure}[ht]
\centering
\includegraphics[width=0.8\textwidth]{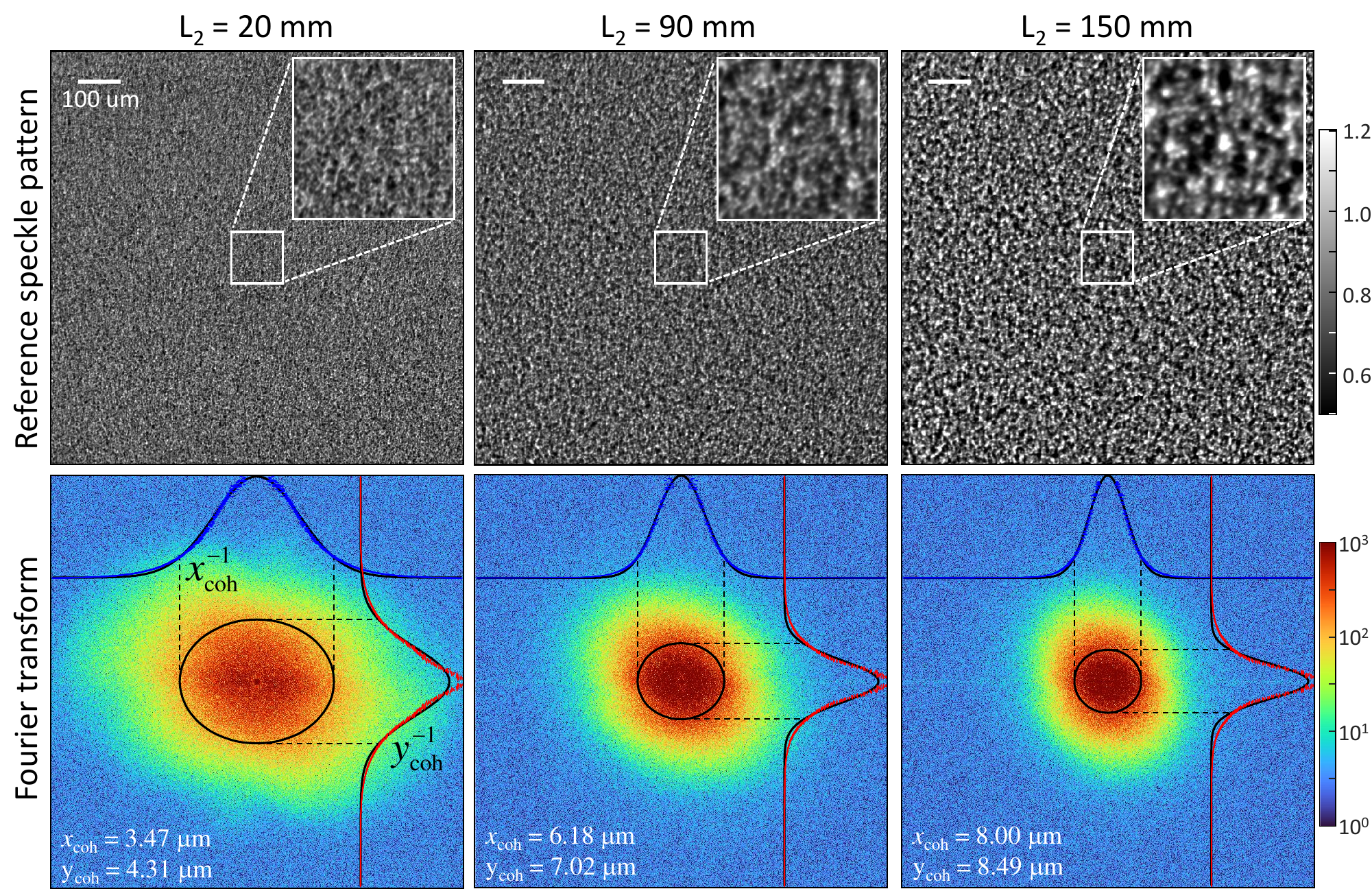}
\caption{
\textbf{Reference speckle patterns and their Fourier transforms} The columns represent different $L_2$ values: \qtylist{20;90;150}{\mm} from left. The top row shows the raw reference speckle patterns, while the bottom row shows their Fourier transforms (i.e., power spectral density). The horizontal and vertical spatial coherence lengths ($x_\mathrm{coh}$ and $y_\mathrm{coh}$) are given in the lower left corner, based on Eq.~\ref{IOTF}.
}\label{figS:speckle}
\end{figure}

\begin{figure}[ht]
\centering
\includegraphics[width=0.5\textwidth]{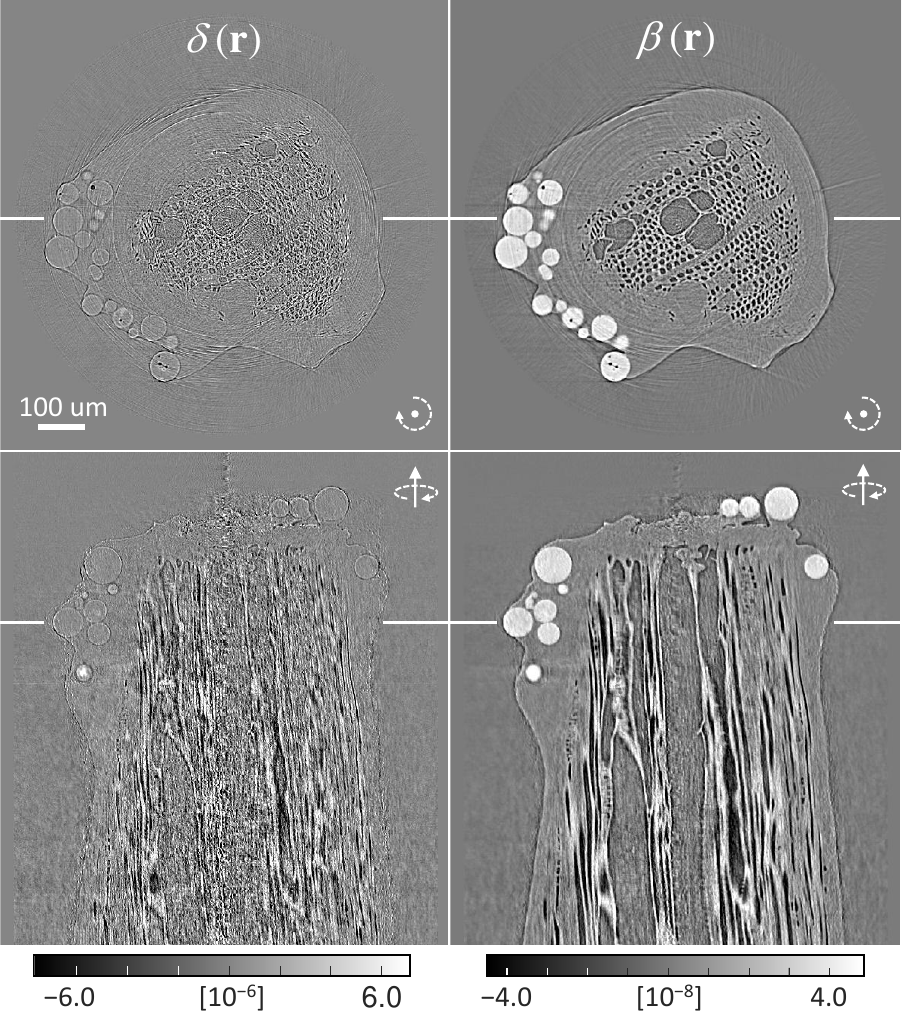}
\caption{
\textbf{Tomographic reconstruction results without the preconditioner}. Without the preconditioner, the algorithm significantly underestimates the phase, resulting in completely incorrect $\delta$ and $\beta$ values. All other parameters are identical to the PWF result shown in Fig.~\ref{fig3}.
}\label{figS:noPrecon}
\end{figure}

\begin{figure}[ht]
\centering
\includegraphics[width=1.0\textwidth]{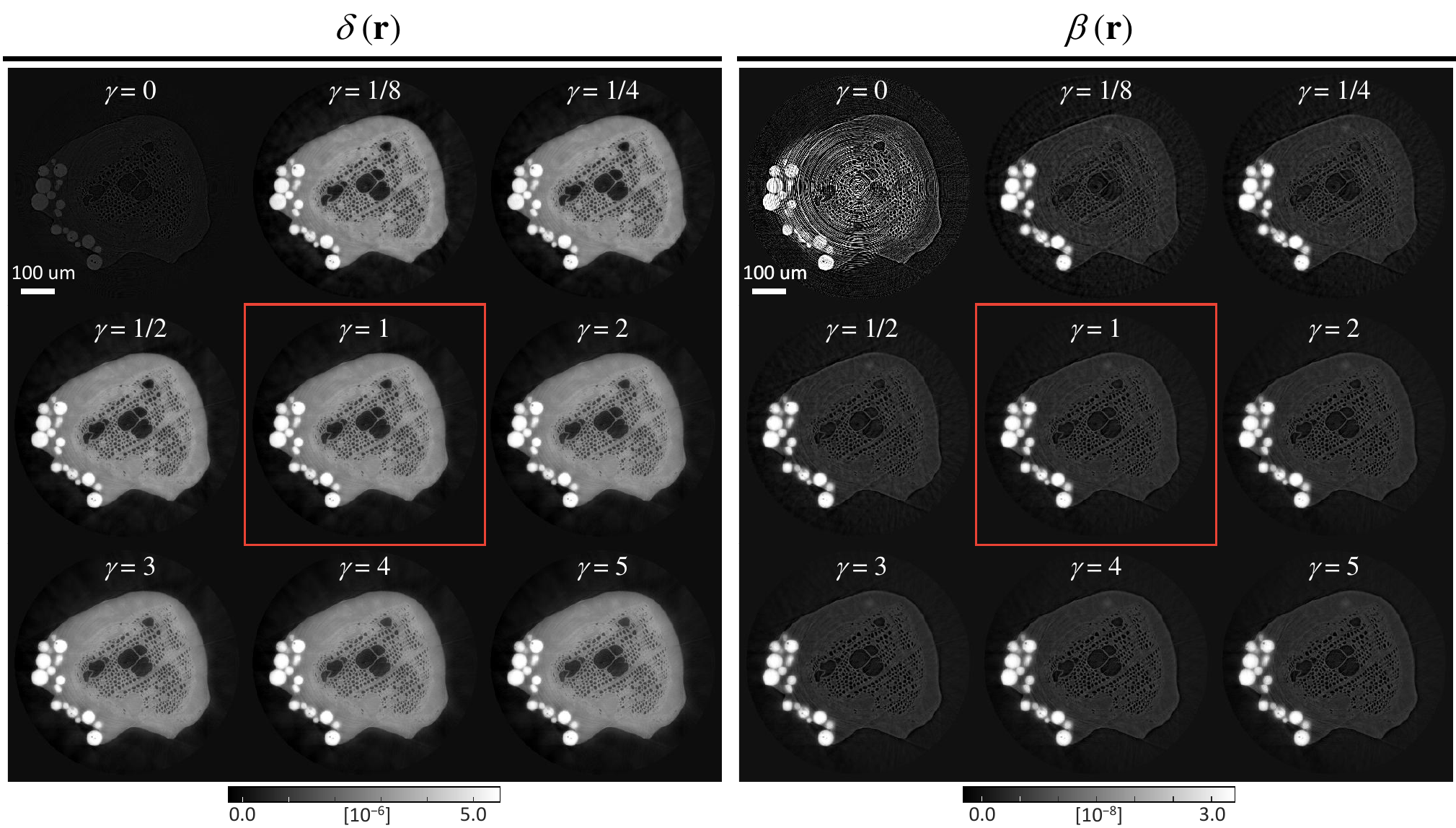}
\caption{
\textbf{Tomographic reconstruction results with different oversampling ratios ($\gamma$).} Different oversampling ratios are applied to the sample transfer function ($\mathrm{STF}_k$, eq.~\ref{eq:STF}) used in the regularization window (Eq.~\ref{eq:regWin}). Subtle changes are observed except for the $\gamma=0$ case, which means no regularization. The $\gamma=1$ is used throughout the paper (red boxes). All other parameters are identical to the PWF result shown in Fig.~\ref{fig3}.
}\label{figS:gamma}
\end{figure}

\begin{figure}[ht]
\centering
\includegraphics[width=1.0\textwidth]{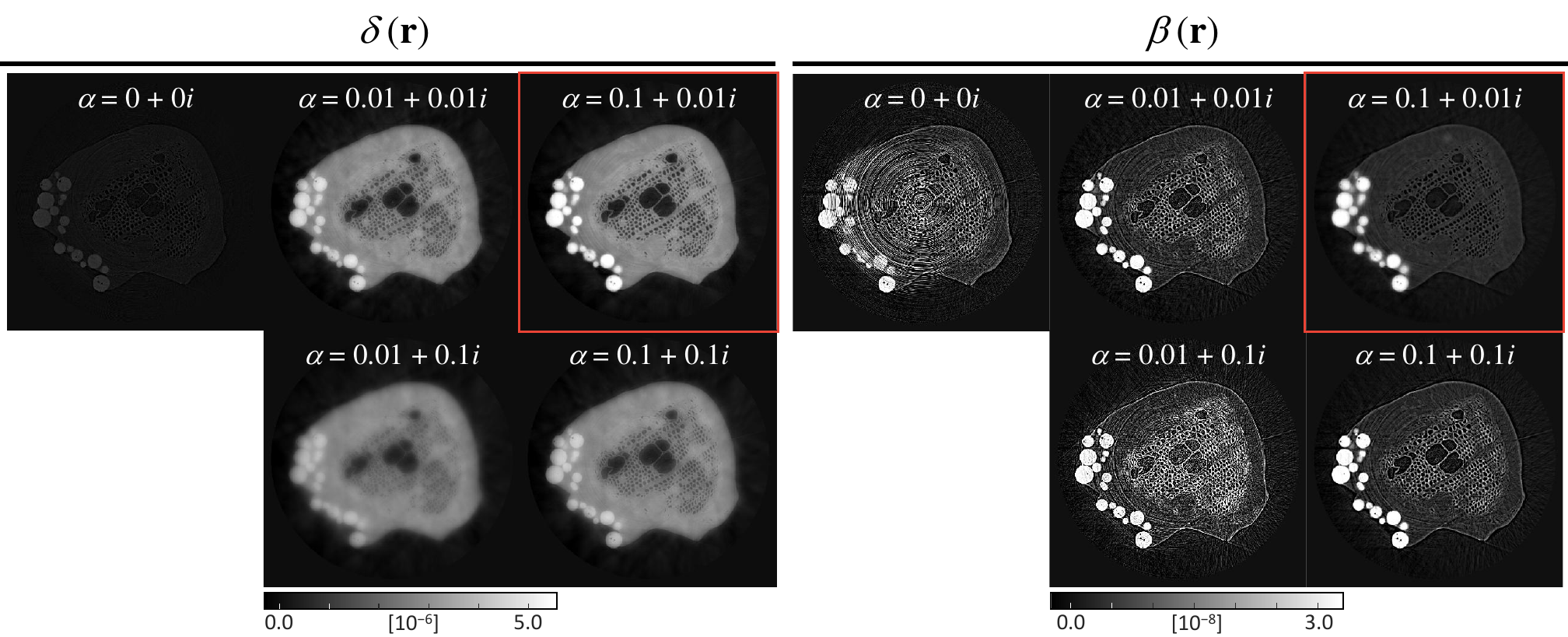}
\caption{
\textbf{Tomographic reconstruction results with different regularization parameters ($\alpha$).} An inappropriate regularization parameter results mainly in additional blur in $\delta(\mathbf{r})$ and edge enhancement in $\delta(\mathbf{r})$. The $\alpha=0$ case means no regularization. The $\alpha=0.1+0.01i$ is used throughout the paper (red boxes). All other parameters are identical to the PWF result shown in Fig.~\ref{fig3}.
}\label{figS:alpha}
\end{figure}

\begin{figure}[ht]
\centering
\includegraphics[width=0.65\textwidth]{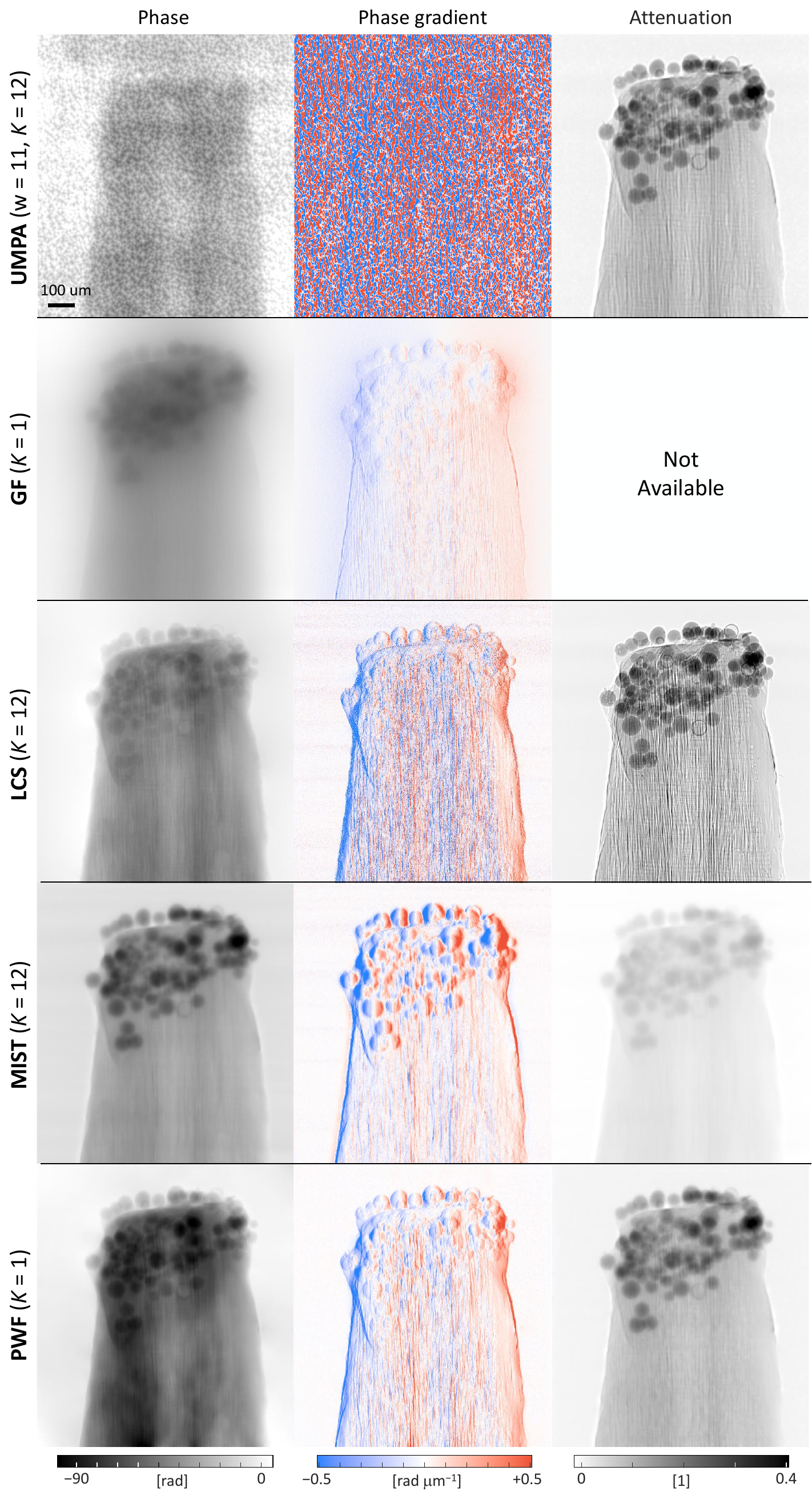}
\caption{
\textbf{Phase retrieval result from different speckle-tracking methods}
\textbf{a}, Unified modulated pattern analysis (UMPA) with $K = 12$ in different window sizes $w= 11$ \cite{zdora2017x}.
\textbf{b}, geometric-flow (GF) speckle tracking with $K = 1$ \cite{paganin2018single}. An attenuation image is not available because GF neglects sample attenuation. 
\textbf{c}, Multimodal intrinsic speckle-tracking (MIST) with $K = 12$ \cite{alloo2023m}. An attenuation image is simply proportional to the phase because MIST assumes that a single material sample has a given ratio between the real and imaginary parts of the refractive index.
\textbf{d}, Low coherence system (LCS) with $K = 12$ \cite{quenot2021implicit}. 
\textbf{e}, Preconditioned Wirtinger flow (flow) with $K = 1$.
The same color scales as in Fig.~\ref{fig2} are used for direct comparison.
}\label{figS:STMcomparison}
\end{figure}

\begin{figure}[ht]
\centering
\includegraphics[width=1.0\textwidth]{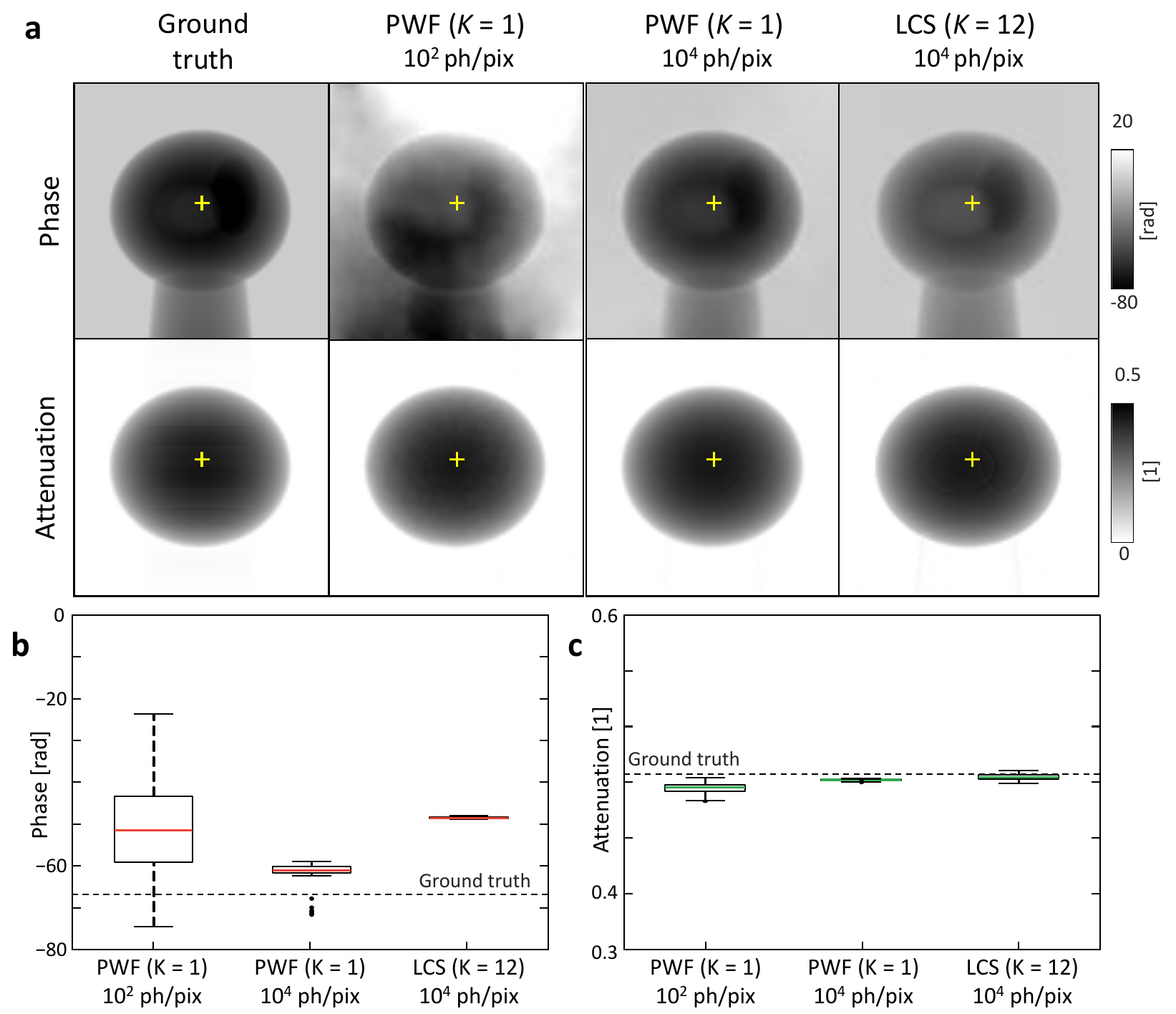}
\caption{
\textbf{Numerically simulated phase reconstruction results of preconditioned Wirtinger flow (PWF) compared with low coherence system (LCS)}
\textbf{a}, We performed numerical experiments using a phantom sample. Assuming a finite number of impinging photons per pixel, we add Poisson noise without a sample or diffuser (ph/pix, shown above). For LCS, we repeated the process 12 times with different diffuser functions. Based on these measurements, we reconstructed the phase attenuation and the images, which are shown in the top and bottom rows, respectively.
\textbf{b} and \textbf{c}, The box plots show the 40 reconstructed phase (b) and attenuation (c) values at the yellow-crossed positions Fig.~\ref{figS:PWFsimul_phase}a. Each measurement used different diffuser functions while maintaining the same phantom sample.
}\label{figS:PWFsimul_phase}
\end{figure}

\begin{figure}[ht]
\centering
\includegraphics[width=1.0\textwidth]{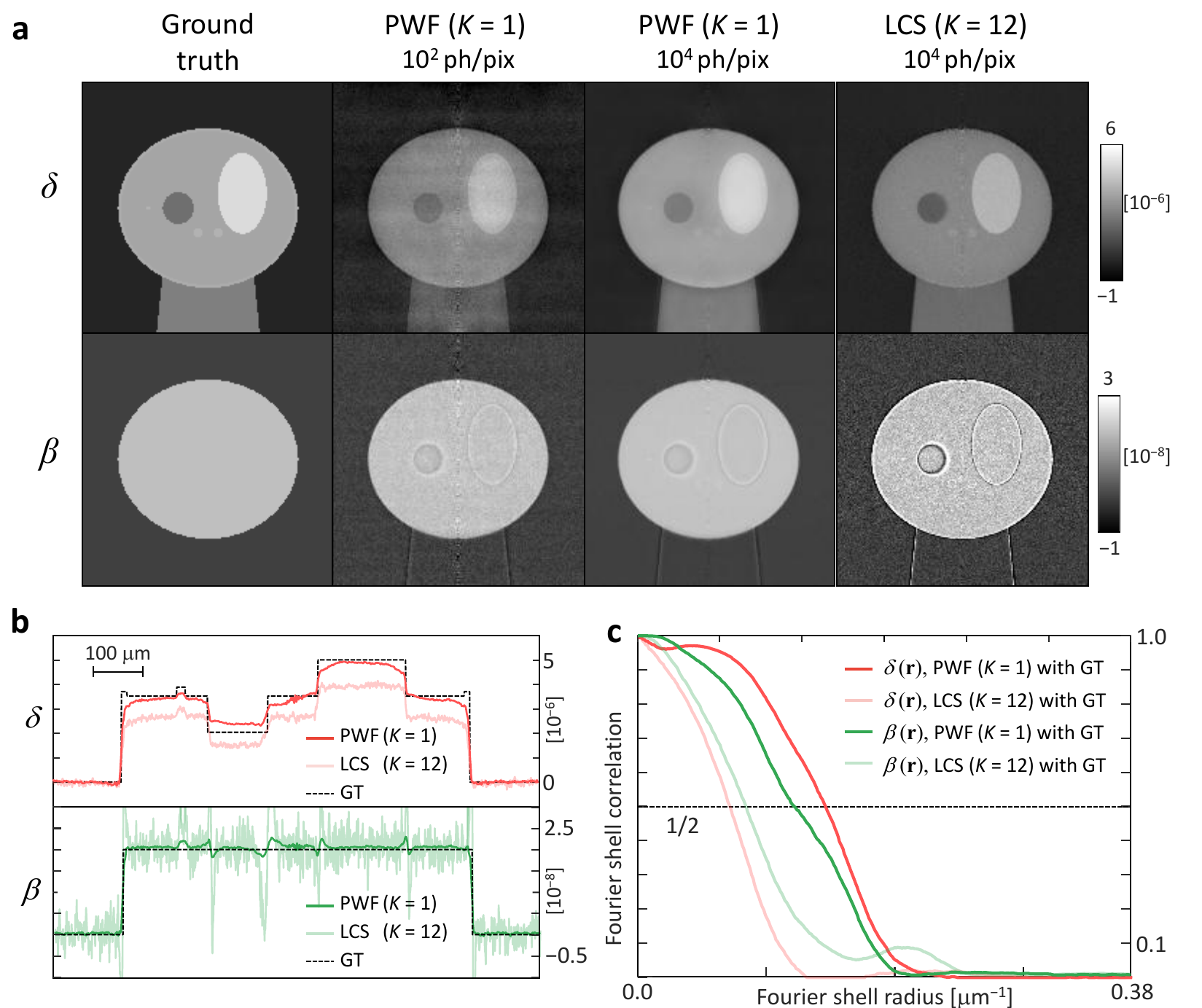}
\caption{
\textbf{Numerically simulated tomogram reconstruction results of preconditioned Wirtinger flow (PWF) compared with low coherence system (LCS)}
\textbf{a},Building on what we did in Fig.~\ref{figS:PWFsimul_phase}, we acquire speckle images at different projection angles by numerically rotating the sample. Based on these measurements, we reconstructed the 3D refractive indices.
\textbf{b}, The line profiles of the simulated results are similar to those in Fig.~\ref{fig3}e. The PWF and LCS profiles are both from the simulated results with $10^4$ ph/pix. The ground truth (GT) profile is shown as dotted lines.
\textbf{c}, The Fourier shell correlation (FSC) of the simulated tomograms with the GT. Since the GT is well known, we can calculate the FSC directly with GT, which is not possible in experimental situations. The 1/2 criterion (dotted line) is shown as the resolution criterion proposed in Eq.~\ref{eq:C=0.5}.
}\label{figS:PWFsimul_tomo}
\end{figure}

\begin{figure}[ht]
\centering
\includegraphics[width=1.0\textwidth]{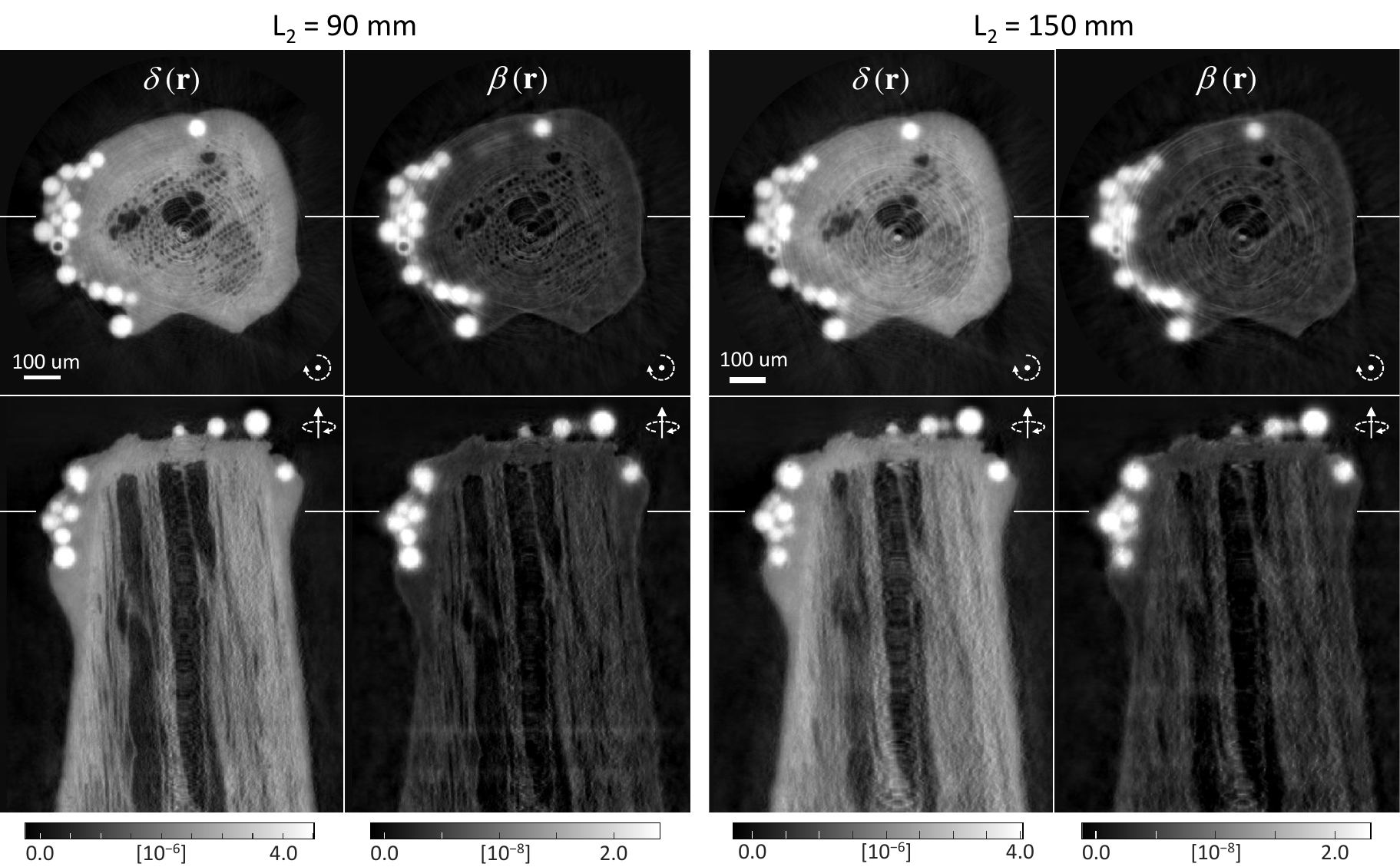}
\caption{
\textbf{Tomographic reconstruction results in different $L_2$ = \qtylist{90;150}{\mm}}. Due to the increased speckle grain (Fig.~\ref{figS:speckle}), a decrease in spatial resolution is observed as $L_2$ increases.
}\label{figS:L2}
\end{figure}

\begin{figure}[ht]
\centering
\includegraphics[width=1.0\textwidth]{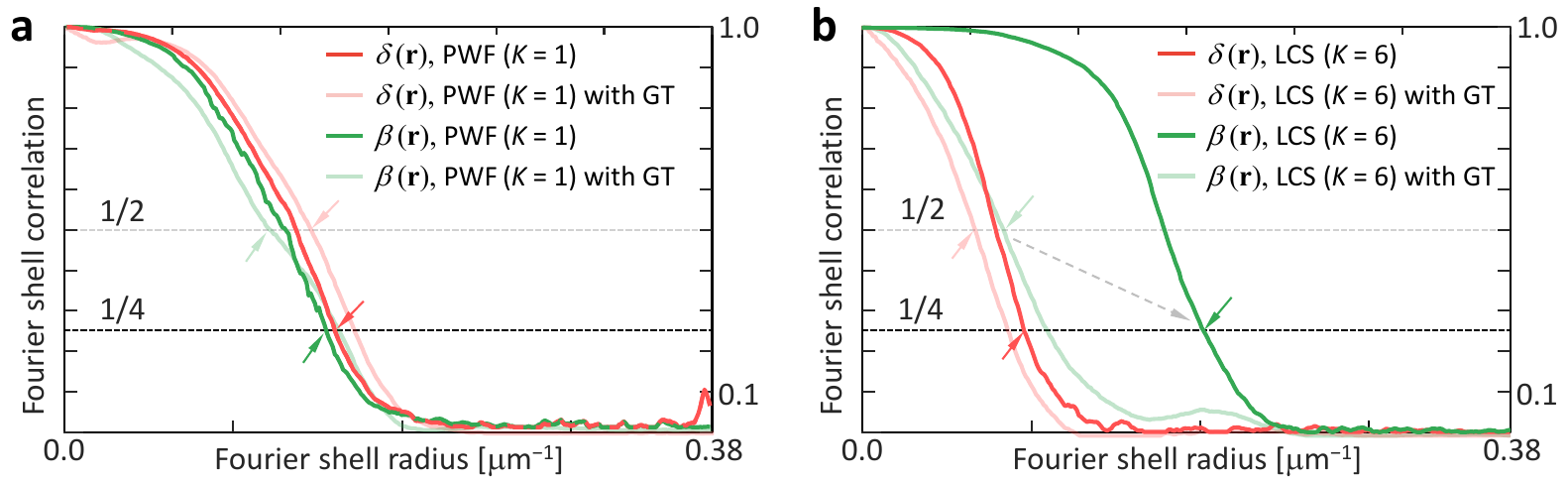}
\caption{
\textbf{Numerical simulation results of Fourier shell correlation (FSC)}
\textbf{a} and \textbf{b}, The numerical FSC results for PWF (a) and LCS (b). Two different FSCs are presented here: the FSC between independent numerical reconstructions (darker colors) and the FSC between numerical reconstructions and the ground truth (GT) (lighter colors). The former simulates the experimental FSC calculated in Fig.~\ref{fig3}f, and the latter provides the actual resolution of the reconstructed tomogram. The resolution criteria of the two FSCs are $1/4$ and $1/2$, respectively (see Methods). The estimated and actual resolutions are indicated as darker and lighter arrows, respectively. A significant discrepancy in $\beta(\mathbf{r})$ of LCS is highlighted by a dotted gray arrow.
}\label{figS:FSCsimul}
\end{figure}

\begin{figure}[ht]
\centering
\includegraphics[width=1.0\textwidth]{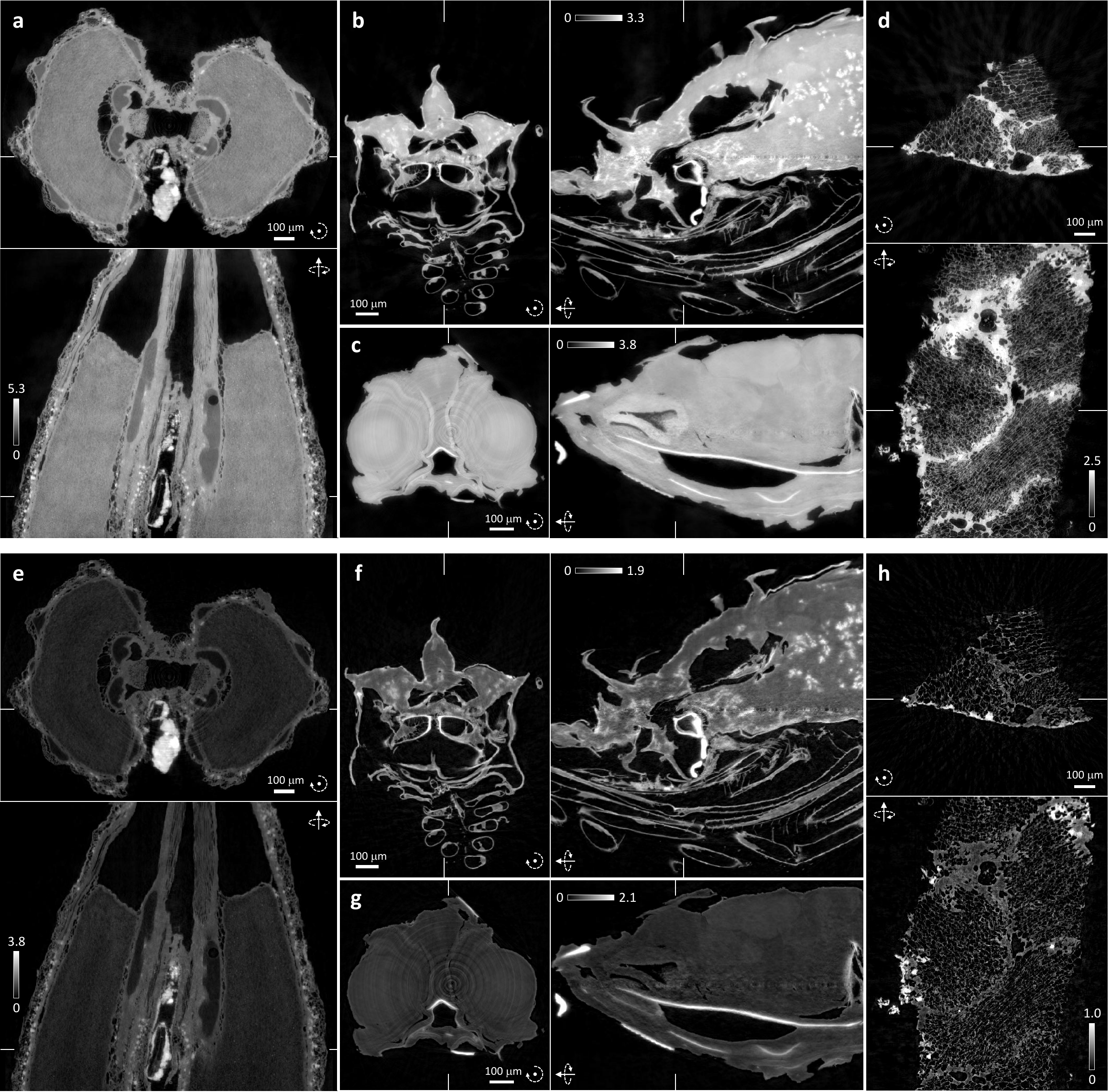}
\caption{
\textbf{The reconstructed complex refractive index---$\delta(\mathbf{r})$ and $\beta(\mathbf{r})$---of the samples in Fig.~\ref{fig4}.}
\textbf{a} and \textbf{e}, A cumin seed; \textbf{b} and \textbf{f}, a dried shrimp; \textbf{c} and \textbf{g}, a dried anchovy; and \textbf{d} and \textbf{h}, a piece of cork, where \textbf{a--d} and \textbf{e--h} depicts $\delta(\mathbf{r})$ and $\beta(\mathbf{r})$, respectively. 
The $\delta(\mathbf{r})$ results are identical to Fig.~\ref{fig4}, but repeated here for easier comparison. The colorbar units are $10^{-6}$ and $10^{-8}$ for \textbf{a--d} and \textbf{e--h}, respectively.
}\label{figS:fig4atten}
\end{figure}

\begin{figure}[ht]
\centering
\includegraphics[width=1.0\textwidth]{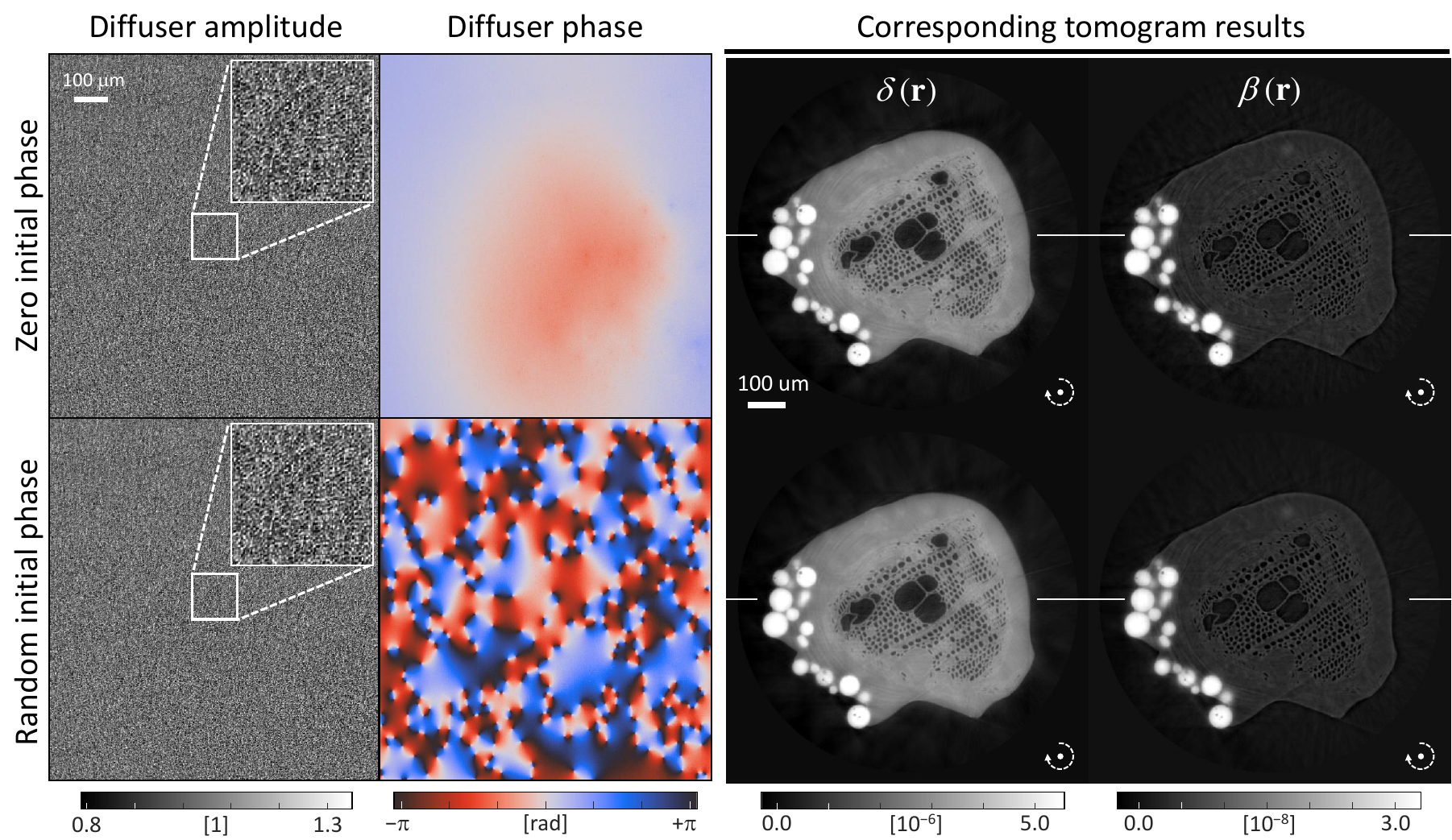}
\caption{
\textbf{Simultaneous reconstruction of the diffuser transmission function ($t_r$) and the corresponding tomogram results.} As indicated above, the columns represent the reconstructed diffuser amplitude ($|t_r|$), reconstructed diffuser angle ($\angle t_r$), and the corresponding tomogram results.
The phase of the reference speckle is initialized to zero (top row) or random (bottom row). Since all the projection angles share the same diffuser, $t_r$ is continuously updated across all the 801 projection angles. Despite significant differences in the diffuser phase, the amplitude part remains almost identical. The diffuser phase does not significantly affect the sample field retrieval, resulting in tomogram results that are almost identical to those in Fig.~\ref{fig3}, where the phase of the reference speckle was set to zero throughout the field retrieval sequence. All parameters are identical to the PWF result shown in Fig.~\ref{fig3}. For the tomogram results, the same color scales as in Fig.~\ref{fig2} are used for direct comparison.
}\label{figS:diffPhase}
\end{figure}

\clearpage
\noindent \textbf{Video S1} \quad \textbf{Cross sections of the toothpick with glass beads.}
These are cross sections of the reconstructed tomogram, one perpendicular (left) and one parallel (right) to the rotation (Y-)axis. The left panel shows the cross section from top to bottom, while the right panel shows the cross section from back to front. The intersection of the two cross sections is shown as a yellow vertical line in both panels. The scale bar indicates \qty{100}{\um}

\bigskip

\noindent \textbf{Video S2} \quad \textbf{Cross sections of a cumin seed.}
These are cross sections of the reconstructed tomogram, one perpendicular (left) and one parallel (right) to the rotation (Y-)axis. The left panel shows the cross section from top to bottom, while the right panel shows the cross section from back to front. The intersection of the two cross sections is shown as a yellow vertical line in both panels. The scale bar indicates \qty{100}{\um}

\bigskip

\noindent \textbf{Video S3} \quad \textbf{Cross sections of  a dried shrimp.}
These are cross sections of the reconstructed tomogram, one perpendicular (left) and one parallel (right) to the rotation (Y-)axis. The left panel shows the cross section from top to bottom, while the right panel shows the cross section from back to front. The intersection of the two cross sections is shown as a yellow vertical line in both panels. The scale bar indicates \qty{100}{\um}

\bigskip

\noindent \textbf{Video S4} \quad \textbf{Cross sections of a dried anchovy.}
These are cross sections of the reconstructed tomogram, one perpendicular (left) and one parallel (right) to the rotation (Y-)axis. The left panel shows the cross section from top to bottom, while the right panel shows the cross section from back to front. The intersection of the two cross sections is shown as a yellow vertical line in both panels. The scale bar indicates \qty{100}{\um}
\bigskip

\noindent \textbf{Video S5} \quad \textbf{Cross sections of a piece of cork.}
These are cross sections of the reconstructed tomogram, one perpendicular (left) and one parallel (right) to the rotation (Y-)axis. The left panel shows the cross section from top to bottom, while the right panel shows the cross section from back to front. The intersection of the two cross sections is shown as a yellow vertical line in both panels. The scale bar indicates \qty{100}{\um}

\bigskip

\noindent \textbf{Video S6} \quad \textbf{Experimental data measurement.}
Raw sample speckle measurement video performed at the 6C beamline of PLS-II in Korea.



\end{document}